\documentclass[10pt]{article}

\usepackage{srcltx}

\usepackage{amsmath}
\usepackage{amssymb,amsfonts}

\usepackage{psfrag}
\usepackage{enumerate}
\usepackage{subfigure}
\usepackage{epsfig}
\usepackage{color}
\usepackage[usenames,dvipsnames]{xcolor}

\usepackage{booktabs}

\usepackage{fullpage}
\usepackage{setspace}
\usepackage{flushend}
\usepackage{multicol}
\usepackage{multirow}

\usepackage{cite}
\usepackage{url}

\newtheorem{definition}{Definition}

\makeatletter

\newenvironment{rsmallmatrix}{\null\,\vcenter\bgroup
  \Let@\restore@math@cr\default@tag
  \baselineskip6\ex@ \lineskip1.5\ex@ \lineskiplimit\lineskip
  \ialign\bgroup\hfil$\m@th\scriptstyle##$&&\thickspace\hfil
  $\m@th\scriptstyle##$\crcr
}{%
  \crcr\egroup\egroup\,%
}
\makeatother

\title{%
Low-complexity 8-point DCT Approximations \\
Based on Integer Functions}

\author{%
R.~J.~Cintra%
\thanks{%
R.~J.~Cintra is with the
Signal Processing Group,
Departamento de Estat\'istica, 
Universidade Federal de Pernambuco,
Recife, Brazil.
E-mail:~\protect\url{rjdsc@dsp.ufpe.org}}
\and
F.~M.~Bayer%
\thanks{F.~M.~Bayer 
is with the 
Departamento de Estat\'istica 
and
Laborat\'orio de Ci\^encias Espaciais de Santa Maria (LACESM), 
Universidade Federal de Santa Maria, Santa Maria, RS, Brazil.
Email:~\protect\url{bayer@ufsm.br}.}
\and
C.~J.~Tablada%
\thanks{C.~J.~Tablada is with the 
Signal Processing Group
and
the Graduate Program in Statistics,
Universidade Federal de Pernambuco, Recife, PE, Brazil.}
}

\date{}

\begin{document}

\maketitle

\onehalfspacing

\begin{abstract}
In this paper,
we propose a collection
of approximations
for the 8-point discrete cosine transform (DCT)
based on integer functions.
Approximations could be systematically obtained
and several existing approximations
were identified as particular cases.
Obtained approximations were compared with the DCT and
assessed in the context of JPEG-like image compression.
\end{abstract}

\section{Introduction}

The discrete cosine transform (DCT) 
is widely regarded
as a key operation
in 
digital signal processing~\cite{rao1990discrete, britanak2007discrete}.
In fact,
the
Karhunen-Lo\`eve transform~(KLT)
is the asymptotic
equivalent
of the DCT,
being the former
an optimal transform 
in terms
of decorrelation and energy compaction
properties~\cite{ahmed1974, Clarke1981, rao1990discrete, britanak2007discrete, Liang2001, haweel2001new}.
When
high correlated
first-order Markov signals are considered~\cite{rao1990discrete, britanak2007discrete}---such as natural images~\cite{Liang2001}---
the DCT can closely
emulate the KLT~\cite{ahmed1974}.

The DCT has been considered and effectively adopted
in a number of methods for image and video coding~\cite{bhaskaran1997}.
In fact,
the DCT is the central mathematical operation
for
the following standards:
JPEG~\cite{Wallace1992,penn1992}, 
MPEG-1~\cite{roma2007hybrid}, 
MPEG-2~\cite{mpeg2}, 
H.261~\cite{h261}, 
H.263~\cite{h263}, 
H.264~\cite{wiegand2003,h264, h2642003, wiegand2003},
and 
the recent HEVC~\cite{hevc1, Bossen2012, Sullivan2012}.
In all above standards,
the particular 8-point DCT is considered.

Thus,
developing fast algorithms
for the efficient evaluation
of the 8-point DCT is a main task
in the circuits, systems, and signal processing communities.
Archived literature
contains a multitude of fast algorithms 
for this particular blocklength~\cite{vetterli1984simple, hou1987fast}.
Remarkably
extensive reports have been generated
amalgamating scattered results 
for the 8-point DCT~\cite{rao1990discrete,britanak2007discrete}.
Among the most
popular techniques,
we mention
the following algorithms:
Wang factorization~\cite{wang1984fast},
Lee DCT for power-of-two blocklengths~\cite{lee1984new},
Arai DCT scheme~\cite{arai1988fast},
Loeffler algorithm~\cite{loeffler1991practical},
Vetterli-Nussbaumer algorithm~\cite{vetterli1984simple}, 
Hou algorithm~\cite{hou1987fast},
and
Feig-Winograd factorization~\cite{fw1992}.
All these methods are classical results in the field
and
have been considered 
for practical applications~\cite{roma2007hybrid, Vasudev1998, Lin2006}.
For instance,
the Arai DCT scheme was employed in various recent
hardware implementations of the DCT~\cite{madanayake2011, rajapaksha2013asynchronous, edirisuriya2012vlsi}.

Naturally,
DCT fast algorithms that result in major computational savings
compared to direct computation of the DCT were already developed decades ago.
In fact,
the intense research in the field
has led to
methods 
that 
are very close to 
the theoretical
complexity
of DCT~\cite{Liang2001,
Madanayake2013, 
Heideman1988,
loeffler1991practical,
arai1988fast}.
Thus,
the computation of the exact DCT
is a task with very little room for major improvements
in terms of minimization of computational complexity
by means of standard methods.

On the other hand,
DCT approximations---operations that closely emulate the DCT---
are
mathematical tools
that can furnish an alternative venue for
DCT evaluation.
Effectively,
DCT approximations
have already been considered in a number of works~\cite{britanak2007discrete,
haweel2001new,
bc2012,
bouguezel2008multiplication,
bas2011}.
Moreover,
although usual fast
algorithms can 
reduce the computational complexity significantly, 
they still need floating-point operations~\cite{britanak2007discrete}. 
In contrast,
approximate DCT methods
can be tailored
to require very low arithmetic complexity.

A comprehensive list of approximate methods for the DCT 
is found in~\cite{britanak2007discrete}.
Prominent techniques include
the
signed DCT (SDCT)~\cite{haweel2001new},
the 
binDCT~\cite{Liang2001},
the
level 1 approximation by 
Lengwehasatit-Ortega~\cite{lengwehasatit2004scalable},
the
Bouguezel-Ahmad-Swamy (BAS) series of algorithms~\cite{bas2008, bouguezel2008multiplication, bas2009, bas2010, bas2011, bas2013},
the
DCT round-off approximation~\cite{cb2011},
the 
modified DCT round-off approximation~\cite{bc2012},
and
the
multiplier-free DCT approximation for RF imaging~\cite{multibeam2012}.

The goal of this paper is two-fold.
First,
we aim
at proposing a 
systematic procedure 
for deriving 
low-complexity approximations
for the 8-point DCT.
For such,
we consider
several 
types of rounding-off functions
applied to 
scaled the exact 8-point DCT matrix.
The entries of 
the sough approximate DCT matrices
are required to
possess null multiplicative complexity;
only additions and simple bit-shifting operations
are allowed.
Second,
we focus on suggesting practical approximations
and
assessing them
as tools
for JPEG-like image compression.

The paper unfolds as follows. 
In Section~\ref{sec:DCT_Approximations},
we describe the mathematical framework 
of the DCT 
and 
we discuss
the polar decomposition method for
DCT approximation~\cite{cintra2011integer}.
In Section~\ref{sec:Scaling_and_Integer_Mapping},
we propose a systematic method based on integer functions
for obtaining
low-complexity matrices
useful for generating DCT approximations.
The discussed method is based on
a computational search
over a subset of candidate matrices
under
constraints
of
low-computational complexity 
and orthogonality or near orthogonality.
In Section~\ref{sec:Computational_Search},
we provide details of the computational search
and list the obtained approximations.
Section~\ref{sec:Fast_Algorithm} presents fast algorithms
for the obtained approximations. 
In Section~\ref{sec:Assessment}, 
the resulting approximations are subject to performance assessment 
in the context of image compression
using image quality measures as figures of merit. 
In Section~\ref{sec:Conclusion},
we state concluding remarks.

\section{Exact and Approximate DCT}
\label{sec:DCT_Approximations}

\subsection{Mathematical Preliminaries}

The $N$-point DCT is algebraically represented by
the $N \times N$ transformation matrix 
$\mathbf{C}_N$
whose elements are given by~\cite{rao1990discrete, britanak2007discrete} 
\begin{align*}
c_{m,n}=
\frac{1}{\sqrt{N}}
\,
\beta_{m-1}
\cos
\left(
\frac{\pi (m-1)(2n-1)}{2N} 
\right)
, 
\end{align*} 
where $m,n=1,2,\ldots,N$,
$\beta_0 = 1$,
and
$\beta_k = \sqrt{2}$, for $k\neq0$.
Let
$\mathbf{x}=\begin{bmatrix} x_0 & x_1 & \cdots & x_{N-1} \end{bmatrix}^\top$
be an input vector,
where the superscript ${}^\top$ denotes the transposition operation.
The one-dimensional (\mbox{1-D}) DCT transform of~$\mathbf{x}$
is 
the $N$-point vector~$\mathbf{X}=\begin{bmatrix} X_0 & X_1 & \cdots & X_{N-1} \end{bmatrix}^\top$
given by
$\mathbf{X}=\mathbf{C}_N \cdot \mathbf{x}$.
Because $\mathbf{C}_N$ is an orthogonal matrix,
the inverse transformation can be written according to 
$\mathbf{x}=\mathbf{C}_N^{\top} \cdot \mathbf{X}$.

Let $\mathbf{A}$ and $\mathbf{B}$ be square matrices
of size~$N$.
For two-dimensional (\mbox{2-D}) signals,
we have the following expressions that relate the forward and inverse 2-D DCT operations, respectively:
\begin{align}
\label{equation-2d-transformation}
\mathbf{B} 
=
\mathbf{C}_N \cdot \mathbf{A} \cdot \mathbf{C}_N^{\top}
\quad
\text{and}
\quad
\mathbf{A} 
=
\mathbf{C}_N ^{\top} \cdot \mathbf{B} \cdot \mathbf{C}_N
. 
\end{align}

Although the procedures described in this work
can be applied to any blocklength,
we focus exclusively on the 8-point DCT.
Thus, 
for simplicity,
the 8-point DCT matrix 
is denoted as~$\mathbf{C}$ and is given by:

\begin{align*}
&\mathbf{C}
\triangleq
\mathbf{C}_8
=
\frac{1}{2}
\cdot
\begin{bmatrix}
\begin{rsmallmatrix}
\gamma_3 & \gamma_3 & \gamma_3 & \gamma_3 & \gamma_3 & \gamma_3 & \gamma_3 & \gamma_3 \\
\gamma_0 & \gamma_2 & \gamma_4 & \gamma_6 & -\gamma_6 & -\gamma_4 & -\gamma_2 & -\gamma_0 \\
\gamma_1 & \gamma_5 & -\gamma_5 & -\gamma_1 & -\gamma_1 & -\gamma_5 & \gamma_5 & \gamma_1 \\
\gamma_2 & -\gamma_6 & -\gamma_0 & -\gamma_4 & \gamma_4 & \gamma_0 & \gamma_6 & -\gamma_2 \\
\gamma_3 & -\gamma_3 & -\gamma_3 & \gamma_3 & \gamma_3 & -\gamma_3 & -\gamma_3 & \gamma_3 \\
\gamma_4 & -\gamma_0 & \gamma_6 & \gamma_2 & -\gamma_2 & -\gamma_6 & \gamma_0 & -\gamma_4 \\
\gamma_5 & -\gamma_1 & \gamma_1 & -\gamma_5 & -\gamma_5 & \gamma_1 & -\gamma_1 & \gamma_5 \\
\gamma_6 & -\gamma_4 & \gamma_2 & -\gamma_0 & \gamma_0 & -\gamma_2 & \gamma_4 & -\gamma_6
\end{rsmallmatrix}
\end{bmatrix}
,
\end{align*}
where
$\gamma_k = \cos(2\pi (k+1) /32)$,
$k=0,1,\ldots,6$.
These quantities are algebraic integers explicitly given by~\cite{madanayake2011}:
\begin{align*}
\gamma_0 &= \frac{\sqrt{2 + \sqrt{2+\sqrt{2}}}}{2} \approx 0.9808\ldots,
\qquad
\gamma_1 = \frac{\sqrt{2+\sqrt{2}}}{2} \approx 0.9239\ldots,
\\
\gamma_2 &= \frac{\sqrt{2 + \sqrt{2-\sqrt{2}}}}{2}\approx 0.8315\ldots,
\qquad
\gamma_3 = \frac{\sqrt{2}}{2}\approx 0.7071\ldots,
\\
\gamma_4 &= \frac{\sqrt{2 - \sqrt{2-\sqrt{2}}}}{2}\approx 0.5556\ldots,
\qquad
\gamma_5 = \frac{\sqrt{2-\sqrt{2}}}{2}\approx 0.3827\ldots,
\\
\gamma_6 &= \frac{\sqrt{2 - \sqrt{2+\sqrt{2}}}}{2}\approx 0.1951\ldots
\end{align*}

In this work,
we adopt the following terminology.
A matrix $\mathbf{A}$ is orthogonal if
$\mathbf{A}\cdot\mathbf{A}^\top$
is a diagonal matrix.
In particular,
if 
$\mathbf{A}\cdot\mathbf{A}^\top$ is the identity matrix,
then $\mathbf{A}$ is said to be orthonormal.

\subsection{DCT Approximations}

Generally,
a DCT approximation
is a transformation~$\hat{\mathbf{C}}$
that---according to some specified metric---
behaves
similarly
to the \emph{exact}
DCT matrix~$\mathbf{C}$.
An approximation matrix~$\hat{\mathbf{C}}$
is usually based on
a transformation matrix~$\mathbf{T}$
of low computational complexity.
Indeed,
matrix~$\mathbf{T}$
is the key component of a given DCT approximation.

Often
the elements of
the transformation matrix~$\mathbf{T}$
possess null multiplicative complexity.
For instance,
this property
can be satisfied
by restricting
the entries of $\mathbf{T}$
to the set of powers of two
$\{0, \pm1, \pm2, \pm4, \pm8, \ldots \}$.
In fact,
multiplications by such elements
are trivial
and require
only bit-shifting operations.

Approximations for the DCT can be classified into
two categories
depending on whether~$\hat{\mathbf{C}}$
is orthonormal or not.
In principle,
given a low-complexity matrix~$\mathbf{T}$
it is possible to derive an orthonormal matrix~$\hat{\mathbf{C}}$
based on~$\mathbf{T}$ by means of
the polar decomposition~\cite{cintra2011integer,seber2008matrix}.
Indeed,
if $\mathbf{T}$ is a full rank real matrix,
then
the following factorization
is uniquely determined:
\begin{align}
\label{eq:approximation}
\hat{\mathbf{C}}
=
\mathbf{S}
\cdot
\mathbf{T}
,
\end{align}
where
$\mathbf{S}$ is 
a symmetric positive definite matrix~\cite[p.~348]{seber2008matrix}.
Matrix~$\mathbf{S}$ is explicitly related to~$\mathbf{T}$
according to the following relation:
\begin{align*}
\mathbf{S} 
=
\sqrt{ (\mathbf{T} \cdot \mathbf{T}^\top)^{-1}}
,
\end{align*}
where
$\sqrt{\cdot}$ denotes
the matrix
square root operation~\cite{Higham1987,MATLAB2013}.
Being orthonormal,
such type of approximation
satisfies
$\hat{\mathbf{C}}^{-1}= \hat{\mathbf{C}}^\top$.
Therefore,
we have that
\begin{align*}
\hat{\mathbf{C}}^{-1} = 
\mathbf{T}^\top \cdot \mathbf{S}
.
\end{align*}
As a consequence,
the inverse transformation
$\hat{\mathbf{C}}^{-1}$
inherits the same computational complexity 
of the forward transformation.

From the computational point of view,
it is desirable that
$\mathbf{S}$ be a diagonal matrix.
In this case,
the computational complexity of
$\hat{\mathbf{C}}$
is the same as that of 
$\mathbf{T}$,
except for the
scale factors in the diagonal matrix~$\mathbf{S}$.
Moreover,
depending on the considered application,
even the constants in~$\mathbf{S}$ 
can be disregarded in terms of computational complexity assessment.
This occurs when the involved constants are trivial multiplicands,
such as the powers of two.
Another more practical
possibility for neglecting the complexity of~$\mathbf{S}$ 
arises when it can be absorbed into other sections of a larger
procedure.
This is the case in JPEG-like compression,
where the quantization step is present~\cite{Wallace1992}.
Thus, 
matrix~$\mathbf{S}$ can be 
incorporated into 
the quantization matrix~\cite{bc2012,bas2011,lengwehasatit2004scalable,bas2008,bas2009,cb2011,bayer201216pt}.
In terms of the inverse transformation,
it is also beneficial that 
$\mathbf{S}$ is diagonal,
because the complexity
of
$\hat{\mathbf{C}}^{-1}$
becomes essentially that of~$\mathbf{T}^\top$.

In order that~$\mathbf{S}$ be a diagonal matrix,
it is sufficient
that~$\mathbf{T}$ satisfies
the orthogonality condition:
\begin{align}
\label{equation-condition}
\mathbf{T} \cdot \mathbf{T}^\top = \mathbf{D},
\end{align}
where
$\mathbf{D}$ 
is a diagonal matrix~\cite{cintra2011integer}.

If~\eqref{equation-condition} is not satisfied,
then~$\mathbf{S}$ is not a diagonal
and
the advantageous properties of the resulting
DCT approximation are in principle lost.
In this case,
the off-diagonal elements
contribute to a computational complexity increase
and 
the absorption of matrix~$\mathbf{S}$ 
cannot be easily done.
However,
at the expense of not providing an orthogonal approximation,
one may consider approximating 
$\mathbf{S}$ itself
by replacing the off-diagonal elements of~$\mathbf{D}$ by zeros.
Thus,
the resulting matrix~$\hat{\mathbf{S}}$
is given by:
\begin{align*}
\hat{\mathbf{S}}
=
\sqrt{[\operatorname{diag}(\mathbf{T}\cdot\mathbf{T}^{\top})]^{-1}}
,
\end{align*}
where
$\operatorname{diag}(\cdot)$
returns a diagonal matrix with 
the diagonal elements of its matrix argument.
Thus, the non-orthogonal approximation is furnished by:
\begin{align*}
\tilde{\mathbf{C}}
=
\hat{\mathbf{S}}
\cdot
\mathbf{T}
.
\end{align*}
Matrix~$\tilde{\mathbf{C}}$
can be a meaningful approximation
if $\hat{\mathbf{S}}$ is,
in some sense,
close to
$\mathbf{S}$;
or, alternatively,
if $\mathbf{T}$ is almost orthogonal.

From the algorithm designing perspective,
proposing
non-orthogonal approximations
may be a less demanding task,
since~\eqref{equation-condition}
is not required to be satisfied.
However,
since 
$\tilde{\mathbf{C}}$
is not orthogonal,
the inverse transformation
must be cautiously examined.
Indeed,
the inverse transformation
does not employ directly 
the low-complexity
matrix~$\mathbf{T}$
and is given by
\begin{align*}
\tilde{\mathbf{C}}^{-1} = 
\mathbf{T}^{-1} \cdot \hat{\mathbf{S}}^{-1}
.
\end{align*}
Even if $\mathbf{T}$
is a low-complexity matrix,
it is not guaranteed
that
$\mathbf{T}^{-1}$
also possesses low computational complexity figures.
Nevertheless,
it is possible to obtain
non-orthogonal approximations
whose both direct and inverse
transformation matrices
have low computational complexity.
Two prominent examples are the
SDCT~\cite{haweel2001new}
and
the BAS approximation described in~\cite{bas2008}.

\section{Scaling and Integer Mapping}
\label{sec:Scaling_and_Integer_Mapping}

Approximations archived in literature
often possess
transformation matrices
with entries defined on the set $\mathcal{C}_0 = \{ 0, \pm1, \pm2 \}$~\cite{bc2012,cb2011,haweel2001new,bas2011,bas2013}.
Thus
such
transformations 
possess null multiplicative complexity,
because the required arithmetic operations
can be implemented 
exclusively by means of additions and bit-shifting operations.
However,
in~\cite{Ishwar2008,nakagaki2007fast},
an
image compression scheme
based on the Tchebichef transform
was advanced for image compression.
This particular method employs
a discrete transfromation
referred to as 
the discrete Tchebichef transform (DTT)~\cite{Ishwar2008}.
The implied
DTT 
matrix possesses entries
defined on $\mathcal{C}_0$
but also
considers the elements $\pm3$.
Multiplications by constants $\pm3$ can be 
implemented by means of one addition and one bit-shifting operation
($3\cdot x = 2\cdot x + x$).
Thus,
as suggested in~\cite{Ishwar2008},
in this work,
we adopt $\mathcal{C} = \mathcal{C}_0 \cup \{ \pm3 \}$
as the domain set of the entries for
the sought DCT approximations.
Nevertheless,
we emphasize that approximations 
with entries $\pm3$
are expected to possess a higher
computational complexity.

In~\cite{haweel2001new}
Haweel
introduced a simple approach
for designing a DCT approximation.
The DCT approximation termed SDCT was defined as follows~\cite{haweel2001new}:
\begin{align*}
\operatorname{sign}(\mathbf{C}),
\end{align*}
where
$\operatorname{sign}(\cdot)$
is the signum function is applied to each entry of~$\mathbf{C}$
and is given by
\begin{align*}
\operatorname{sign}(x)
=
\begin{cases}
+1, & \text{if $x>0$}, \\
0, & \text{if $x=0$}, \\
-1, & \text{if $x<0$}.
\end{cases}
\end{align*}
The SDCT can be regarded 
as a seminal work in the field of DCT approximations.

Additionally,
in~\cite{cintra2011integer,cb2011,bayer2010image}
a low-complexity DCT approximation
was proposed based on the following matrix:
\begin{align*}
\operatorname{round}(2 \cdot \mathbf{C})
,
\end{align*}
where
$\operatorname{round}(\cdot)$ is the entrywise rounding-off function
as implemented in Matlab.
In this work,
we aim at expanding
and generalizing these approximations.

As a venue to design
DCT approximations,
we consider 
integer 
functions~\cite[Cap.~3]{graham2008concrete}.
An integer function
is simply a function whose values are integers.
We aim
at
mapping
the exact entries of the DCT matrix
into integer quantities.
The resulting matrix is sought to
approximate the DCT.
For such end,
we adopt the following
general mapping:
\begin{equation}
\label{equation-prototype}
\begin{split}
\mathbb{R}
&\longrightarrow
\mathcal{M}_8(\mathbb{Z})
\\
\alpha
&\longmapsto
\operatorname{int}
\left(
\alpha
\cdot
\mathbf{C}
\right)
,
\end{split}
\end{equation}
where
$\mathcal{M}_8(\mathbb{Z})$
is the space of 8$\times$8 matrices
over the set of integers~$\mathbb{Z}$,
$\operatorname{int}(\cdot)$
is a prototype integer function~\cite[p.~67]{graham2008concrete}.
Function $\operatorname{int}(\cdot)$ operates
entrywise over its matrix argument.
Parameter $\alpha$ is termed the expansion factor and
scales the exact DCT matrix
allowing a wide range of possible integer mappings~\cite{plonka2004global}.

Particular examples of integer functions
are
the floor,
the ceiling,
the truncation (round towards zero),
and
the round-away-from-zero function.
These functions are defined,
respectively,
as follows:
\begin{align*}
\operatorname{floor}(x)
&
=
\lfloor x \rfloor
=
\max\, \{m\in\mathbb{Z}\mid m\le x\}
,
\\
\operatorname{ceil}(x)
&
=
\lceil x \rceil
=
\min\,\{n\in\mathbb{Z}\mid n\ge x\}
,
\\
\operatorname{trunc}(x)
&=
\operatorname{sign}(x) \cdot \left\lfloor \left\vert x \right\vert \right\rfloor
,
\\
\operatorname{round}_\text{AFZ}(x)
&=
\operatorname{sign}(x) \cdot \left\lceil \left\vert x \right\vert \right\rceil
,
\end{align*}
where 
$\vert \cdot \vert$ returns the absolute value of its argument.

Another particularly useful integer function
is the round to nearest integer 
function~\cite[p.~73]{Oldham2008atlas}.
This function possesses various 
definitions depending on
its behavior
for input arguments whose fractional part is exactly $1/2$.
Thus,
we have the following rounding-off functions:
round-half-up,
round-half-down,
round-half-away-from-zero,
round-half-towards-zero,
round-half-to-even,
and
round-half-to-odd function.
These different nearest integer functions are,
respectively,
given by:
\begin{align*}
\operatorname{round}_\text{HU}(x)
&=
\left\lfloor  x + \frac{1}{2}  \right\rfloor
,
\\
\operatorname{round}_\text{HD}(x)
&=
\left\lceil  x - \frac{1}{2}  \right\rceil
,
\\
\operatorname{round}_\text{HAFZ}(x)
&=
\operatorname{sign}(x) \cdot \left\lfloor \left\vert x \right\vert + \frac{1}{2}\right\rfloor
,
\\
\operatorname{round}_\text{HTZ}(x)
&=
\operatorname{sign}(x) \cdot \left\lceil \left\vert x \right\vert - \frac{1}{2} \right\rceil
,
\\
\operatorname{round}_\text{EVEN}(x)
&=
\begin{cases}
\left\lfloor x - \frac{1}{2} \right\rfloor, 
& \text{if $\frac{2x-1}{4}\in\mathbb{Z}$,} \\
\left\lfloor x + \frac{1}{2} \right\rfloor, & \text{otherwise,} \\
\end{cases}
\\
\operatorname{round}_\text{ODD}(x)
&=
\begin{cases}
\left\lfloor x + \frac{1}{2} \right\rfloor, 
& \text{if $\frac{2x-1}{4}\in\mathbb{Z}$,} \\
\left\lfloor x - \frac{1}{2} \right\rfloor, & \text{otherwise.} \\
\end{cases}
\end{align*}

The round-half-away-from-zero function
is the implementation employed in the
\texttt{round} function in Matlab/Octave.
The international technical standard ISO/IEC/IEEE~60559:2011
recommends
$\operatorname{round}_\text{EVEN}(\cdot)$
as the nearest integer function
of choice~\cite{2011isoiecieee}.
This latter implementation is adopted in 
the scientific computation software Mathematica~\cite{research2013round}.

\section{Computational Search}
\label{sec:Computational_Search}

\subsection{Problem Setup}

In this section,
we exhaustively compute~\eqref{equation-prototype}
for judiciously chosen
values of $\alpha$
such
that the following
conditions about 
$\mathbf{T}=\operatorname{int}(\alpha \cdot \mathbf{C})$ are satisfied:
\begin{enumerate}[(a)]

\item
matrix $\mathbf{T}$
must possess its elements
defined on
$\mathcal{C}$;

\item
$\mathbf{T} \cdot \mathbf{T}^\top$
must be a diagonal matrix or
must exhibit
a small deviation from diagonality
in the sense described in~\cite{Flury1986};

\item
if $\mathbf{T}$ 
is not orthogonal (cf.~\eqref{equation-condition}),
but 
$\mathbf{T} \cdot \mathbf{T}^\top$
is approximately a diagonal matrix,
then
the inverse matrix $\mathbf{T}^{-1}$
must possess
low-complexity
with 
its
elements
defined on
$\mathcal{C}$.

\end{enumerate}

Condition~(a) ensures that the forward
transformation is a low-complexity operation.
Therefore,
in terms of implementation,
it may require simple hardware structures.
If~\eqref{equation-condition} is satisfied,
then
the inverse transformation is guaranteed to have
low computation complexity.
This is
because $\mathbf{T}$ becomes the transpose of itself,
apart from the multiplication by a diagonal matrix.
On the other hand,
if~\eqref{equation-condition} is not satisfied,
then resulting matrices may not be useful 
in contexts that depend on orthogonalization.
Nevertheless,
if
$\mathbf{T} \cdot \mathbf{T}^\top$
is `almost' diagonal,
then
$\mathbf{T}$ can be of interest.
In that case,
one may explicitly check
whether $\mathbf{T}^{-1}$
has low complexity.
Thus, Condition~(c) is considered.

To quantify the deviation from diagonality,
as required in Condition~(b),
we
adopt the
following measure,
called deviation from diagonality.
\begin{definition}
Let $\mathbf{A}$ be a square matrix.
Then its deviation from diagonality is
given by~\cite{Flury1986}:
\begin{align*}
\operatorname{\delta}(\mathbf{A})
=
1
-
\frac{\|\operatorname{diag}(\mathbf{A})\|_\text{F}}
{\|\mathbf{A}\|_\text{F}}
,
\end{align*}
where
$\|\cdot\|_\text{F}$ denotes the Frobenius norm 
for matrices~\cite{seber2008matrix}.
\end{definition}

As a decision criterion,
we adopt the deviation from diagonality exhibited
by the SDCT as the maximum deviation
acceptable for non-orthogonal approximations.
The SDCT was chosen as a reference transformation
because 
(i)~it has proven good properties~\cite{haweel2001new}
and
(ii)~it is widely employed
in performance comparisons~\cite{cb2011,bc2012,bouguezel2008multiplication,bas2008,bas2009,bas2010,bas2013}.
Thus,
according to this criterion,
Condition~(b) becomes:
\begin{enumerate}[(a)]

 \setcounter{enumi}{1}

\item
$\operatorname{\delta}
\left(
\mathbf{T} \cdot \mathbf{T}^\top
\right) 
\leq 
1 - \frac{2}{\sqrt{5}} 
\approx 0.1056$.

\end{enumerate}

In order that
the entries 
of
$\operatorname{int}(\alpha \cdot \mathbf{C})$ 
are defined on $\mathcal{C}$,
we must restrict the range of $\alpha$.
We notice 
that the largest element of $\mathbf{C}$ is $\gamma_0/2$.
Thus,
it is sufficient
to solve the following inequality
for $\alpha$:
$0\leq \operatorname{int}(\alpha \cdot \gamma_0 /2) \leq 3$.
For
the ceiling,
floor,
truncation,
round-away-from-zero,
and all nearest integer functions,
respectively,
we have the following ranges of $\alpha$:
$\left[0, \frac{6}{\gamma_0}\right]$,
$\left[\frac{2}{\gamma_0}, \frac{8}{\gamma_0}\right]$,
$\left[\frac{2}{\gamma_0}, \frac{8}{\gamma_0}\right]$,
$\left[0, \frac{6}{\gamma_0}\right]$,
and
$\left[\frac{1}{\gamma_0}, \frac{7}{\gamma_0}\right]$.

\subsection{Obtained Approximations}

In terms of Condition~(a),
the ceiling function
could
supply only one low-complexity non-orthogonal
candidate matrix for 
DCT approximation
as shown in~Table~\ref{table-ceil}.
However,
its deviation from diagonality is
exceedingly high.
In fact,
we have that
$\operatorname{\delta}(\tilde{\mathbf{T}}_{0}) \approx 0.4548$.
For such reason,
this particular approximation will not be further considered
in this paper.
Hereafter,
we only list matrices that satisfy all prescribed conditions.

The floor function
could not furnish any matrix
under the prescribed requirements.
On the other hand,
when considering the truncation function,
five matrices were obtained,
being listed in
Table~\ref{table-truncation}.
Both orthogonal and non-orthogonal matrices
were found.
Similarly,
for the round-away-from-zero function,
a set of
four distinct matrices was derived.
These matrices are shown
in
Table~\ref{table-round-away-from-zero}.
We notice that 
matrices~$\mathbf{T}_4$ and~$\tilde{\mathbf{T}}_{2}$
coincide with 
the rounded DCT reported in~\cite{cb2011}
and
the SDCT~\cite{haweel2001new},
respectively.
Moreover,
we have that
\begin{align}
\label{equation-equivalence}
\tilde{\mathbf{T}}_4
=
\left[
\begin{smallmatrix}
2 & 0 & 0 & 0 & 0 & 0 & 0 & 0 \\
0 & 1 & 0 & 0 & 0 & 0 & 0 & 0 \\
0 & 0 & 1 & 0 & 0 & 0 & 0 & 0 \\
0 & 0 & 0 & 1 & 0 & 0 & 0 & 0 \\
0 & 0 & 0 & 0 & 2 & 0 & 0 & 0 \\
0 & 0 & 0 & 0 & 0 & 1 & 0 & 0 \\
0 & 0 & 0 & 0 & 0 & 0 & 1 & 0 \\
0 & 0 & 0 & 0 & 0 & 0 & 0 & 1 \\
\end{smallmatrix}
\right]
\cdot
\tilde{\mathbf{T}}_3
.
\end{align}
As a consequence,
by means of~\eqref{eq:approximation},
both 
$\tilde{\mathbf{T}}_3$ and $\tilde{\mathbf{T}}_4$
lead to the same DCT approximation.

\begin{table}
\centering
\caption{Ceiling function}
\label{table-ceil}
\begin{tabular}{ccccc}
\toprule
Approximation & 
Transformation Matrix &
Range of $\alpha$ 
&
Orthogonal?
\\
\midrule
$\tilde{\mathbf{T}}_{0}$ 
&
$\left[
\begin{rsmallmatrix}
1 & 1 & 1 & 1 & 1 & 1 & 1 & 1 \\1 & 1 & 1 & 1 & 0 & 0 & 0 & 0 \\1 & 1 & 0 & 0 & 0 & 0 & 1 & 1 \\1 & 0 & 0 & 0 & 1 & 1 & 1 & 0 \\1 & 0 & 0 & 1 & 1 & 0 & 0 & 1 \\1 & 0 & 1 & 1 & 0 & 0 & 1 & 0 \\1 & 0 & 1 & 0 & 0 & 1 & 0 & 1 \\1 & 0 & 1 & 0 & 1 & 0 & 1 & 0 \\
\end{rsmallmatrix}
\right]$
&
$(0, 2/\gamma_4)$
&
No
\\
\bottomrule
\end{tabular}
\end{table}

\begin{table}
\centering
\caption{Truncation function}
\label{table-truncation}
\begin{tabular}{ccccc}
\toprule
Approximation & 
Transformation Matrix &
Range of $\alpha$ 
&
Orthogonal?
\\
\midrule
$\mathbf{T}_{0}$ 
&
$
\left[
\begin{rsmallmatrix}
1 & 1 & 1 & 1 & 1 & 1 & 1 & 1 \\1 & 1 & 1 & 0 & 0 & -1 & -1 & -1 \\1 & 0 & 0 & -1 & -1 & 0 & 0 & 1 \\1 & 0 & -1 & -1 & 1 & 1 & 0 & -1 \\1 & -1 & -1 & 1 & 1 & -1 & -1 & 1 \\1 & -1 & 0 & 1 & -1 & 0 & 1 & -1 \\0 & -1 & 1 & 0 & 0 & 1 & -1 & 0 \\0 & -1 & 1 & -1 & 1 & -1 & 1 & 0
\end{rsmallmatrix}
\right]
$
&
$(2/\gamma_4, 4/\gamma_0)$
&
Yes
\\[25pt]
$\mathbf{T}_{1}$ 
&
$\left[
\begin{rsmallmatrix}
1 & 1 & 1 & 1 & 1 & 1 & 1 & 1 \\2 & 1 & 1 & 0 & 0 & -1 & -1 & -2 \\0 & 1 & -1 & 0 & 0 & -1 & 1 & 0 \\1 & 0 & -2 & -1 & 1 & 2 & 0 & -1 \\1 & -1 & -1 & 1 & 1 & -1 & -1 & 1 \\1 & -2 & 0 & 1 & -1 & 0 & 2 & -1 \\1 & 0 & 0 & -1 & -1 & 0 & 0 & 1 \\0 & -1 & 1 & -2 & 2 & -1 & 1 & 0
\end{rsmallmatrix}
\right]
$
&
$(4/\gamma_0, 4/\gamma_1)$
&
Yes
\\[25pt]
$\mathbf{T}_{2}$ 
&
$\left[
\begin{rsmallmatrix}
1 & 1 & 1 & 1 & 1 & 1 & 1 & 1 \\2 & 1 & 1 & 0 & 0 & -1 & -1 & -2 \\2 & 0 & 0 & -2 & -2 & 0 & 0 & 2 \\1 & 0 & -2 & -1 & 1 & 2 & 0 & -1 \\1 & -1 & -1 & 1 & 1 & -1 & -1 & 1 \\1 & -2 & 0 & 1 & -1 & 0 & 2 & -1 \\0 & -2 & 2 & 0 & 0 & 2 & -2 & 0 \\0 & -1 & 1 & -2 & 2 & -1 & 1 & 0 \\
\end{rsmallmatrix}
\right]$
&
$(4/\gamma_1, 4/\gamma_2)$
&
Yes
\\[25pt]
$\mathbf{T}_{3}$ 
&
$\left[
\begin{rsmallmatrix}
2 & 2 & 2 & 2 & 2 & 2 & 2 & 2 \\3 & 2 & 2 & 0 & 0 & -2 & -2 & -3 \\3 & 1 & -1 & -3 & -3 & -1 & 1 & 3 \\2 & 0 & -3 & -2 & 2 & 3 & 0 & -2 \\2 & -2 & -2 & 2 & 2 & -2 & -2 & 2 \\2 & -3 & 0 & 2 & -2 & 0 & 3 & -2 \\1 & -3 & 3 & -1 & -1 & 3 & -3 & 1 \\0 & -2 & 2 & -3 & 3 & -2 & 2 & 0 \\
\end{rsmallmatrix}
\right]$
&
$(4/\gamma_4, 6/\gamma_2)$
&
Yes
\\[25pt]
$\tilde{\mathbf{T}}_{1}$ 
&
$\left[
\begin{rsmallmatrix}
1 & 1 & 1 & 1 & 1 & 1 & 1 & 1 \\1 & 1 & 0 & 0 & 0 & 0 & -1 & -1 \\1 & 0 & 0 & -1 & -1 & 0 & 0 & 1 \\1 & 0 & -1 & 0 & 0 & 1 & 0 & -1 \\1 & -1 & -1 & 1 & 1 & -1 & -1 & 1 \\0 & -1 & 0 & 1 & -1 & 0 & 1 & 0 \\0 & -1 & 1 & 0 & 0 & 1 & -1 & 0 \\0 & 0 & 1 & -1 & 1 & -1 & 0 & 0 \\
\end{rsmallmatrix}
\right]$
&
$(2/\gamma_3, 2/\gamma_4)$
&
No
\\
\bottomrule
\end{tabular}
\end{table}

\begin{table}
\centering
\caption{Round-away-from-zero function}
\label{table-round-away-from-zero}
\begin{tabular}{ccccc}
\toprule
Approximation & 
Transformation Matrix &
Range of $\alpha$ 
&
Orthogonal?
\\
\midrule
$\mathbf{T}_{4}$~\cite{cb2011} 
&
$\left[
\begin{rsmallmatrix}
1 & 1 & 1 & 1 & 1 & 1 & 1 & 1 \\1 & 1 & 1 & 0 & 0 & -1 & -1 & -1 \\1 & 1 & -1 & -1 & -1 & -1 & 1 & 1 \\1 & 0 & -1 & -1 & 1 & 1 & 0 & -1 \\1 & -1 & -1 & 1 & 1 & -1 & -1 & 1 \\1 & -1 & 0 & 1 & -1 & 0 & 1 & -1 \\1 & -1 & 1 & -1 & -1 & 1 & -1 & 1 \\0 & -1 & 1 & -1 & 1 & -1 & 1 & 0 \\
\end{rsmallmatrix}
\right]$
& 
$1/\gamma_0$
&
Yes
\\ [25pt]
$\tilde{\mathbf{T}}_{2}$~\cite{haweel2001new}
&
$\left[
\begin{rsmallmatrix}
1 & 1 & 1 & 1 & 1 & 1 & 1 & 1 \\1 & 1 & 1 & 1 & -1 & -1 & -1 & -1 \\1 & 1 & -1 & -1 & -1 & -1 & 1 & 1 \\1 & -1 & -1 & -1 & 1 & 1 & 1 & -1 \\1 & -1 & -1 & 1 & 1 & -1 & -1 & 1 \\1 & -1 & 1 & 1 & -1 & -1 & 1 & -1 \\1 & -1 & 1 & -1 & -1 & 1 & -1 & 1 \\1 & -1 & 1 & -1 & 1 & -1 & 1 & -1 \\
\end{rsmallmatrix}
\right]$
&
$(0, 2/\gamma_0)$
&
No
\\[25pt]
$\tilde{\mathbf{T}}_{3}$
&
$\left[
\begin{rsmallmatrix}
1 & 1 & 1 & 1 & 1 & 1 & 1 & 1 \\2 & 2 & 1 & 1 & -1 & -1 & -2 & -2 \\2 & 1 & -1 & -2 & -2 & -1 & 1 & 2 \\2 & -1 & -2 & -1 & 1 & 2 & 1 & -2 \\1 & -1 & -1 & 1 & 1 & -1 & -1 & 1 \\1 & -2 & 1 & 2 & -2 & -1 & 2 & -1 \\1 & -2 & 2 & -1 & -1 & 2 & -2 & 1 \\1 & -1 & 2 & -2 & 2 & -2 & 1 & -1 \\
\end{rsmallmatrix}
\right]$
&
$(2/\gamma_2, 2/\gamma_3]$
&
No
\\[25pt]
$\tilde{\mathbf{T}}_{4}$
&
$\left[
\begin{rsmallmatrix}
2 & 2 & 2 & 2 & 2 & 2 & 2 & 2 \\2 & 2 & 1 & 1 & -1 & -1 & -2 & -2 \\2 & 1 & -1 & -2 & -2 & -1 & 1 & 2 \\2 & -1 & -2 & -1 & 1 & 2 & 1 & -2 \\2 & -2 & -2 & 2 & 2 & -2 & -2 & 2 \\1 & -2 & 1 & 2 & -2 & -1 & 2 & -1 \\1 & -2 & 2 & -1 & -1 & 2 & -2 & 1 \\1 & -1 & 2 & -2 & 2 & -2 & 1 & -1 \\
\end{rsmallmatrix}
\right]$
&
$(2/\gamma_3, 2/\gamma_4)$
&
No
\\
\bottomrule
\end{tabular}
\end{table}

The nearest integer function may result in different approximations
when
the choice of $\alpha$ results
in a matrix $\alpha \cdot \mathbf{C}$
whose entries are possibly half-integers.
The values of $\alpha$ that effect half-integers
are of the form
$l / \gamma_k$,
$k=0,1,\ldots,6$,
where $l\in\mathbb{Z}$.
Apart from these critical points,
the different types of nearest integer functions
behave identically.
By examining these boundary cases,
we could establish intervals for which 
each of discussed 
nearest integer functions
results in 
meaningful DCT approximations.

Table
\ref{table-round-half-up-down}
brings
the obtained
low-complexity matrices
derived from the 
round-half-up
and
round-half-down functions.
Among the resulting matrices,
we identify
$\mathbf{T}_6$
as the approximation described in~\cite{multibeam2012}.
The remaining nearest integer functions
supply no matrix different from the above listed ones.
In Table~\ref{table-nearest-integer-functions},
we show the resulting matrices.

\begin{table}
\centering
\caption{Round-half-up and round-half-down functions}
\label{table-round-half-up-down}
\begin{tabular}{ccccc}
\toprule
Approximation & 
Transformation Matrix &
Range of $\alpha$  
&
Orthogonal?
\\
\midrule
$\mathbf{T}_{0}$ 
& 
See Table~\ref{table-truncation}&
$(1/\gamma_4, 1/\gamma_5)$
&
Yes
\\
$\mathbf{T}_{4}$~\cite{cb2011} 
& 
See Table~\ref{table-round-away-from-zero}&
$(1/\gamma_5, 3/\gamma_0)$
&
Yes
\\
$\mathbf{T}_{5}$ 
& 
$\left[
\begin{rsmallmatrix}
1 & 1 & 1 & 1 & 1 & 1 & 1 & 1 \\2 & 1 & 1 & 0 & 0 & -1 & -1 & -2 \\1 & 1 & -1 & -1 & -1 & -1 & 1 & 1 \\1 & 0 & -2 & -1 & 1 & 2 & 0 & -1 \\1 & -1 & -1 & 1 & 1 & -1 & -1 & 1 \\1 & -2 & 0 & 1 & -1 & 0 & 2 & -1 \\1 & -1 & 1 & -1 & -1 & 1 & -1 & 1 \\0 & -1 & 1 & -2 & 2 & -1 & 1 & 0 \\
\end{rsmallmatrix}
\right]$
&
$(3/\gamma_0, 3/\gamma_1)$
&
Yes
\\[25pt]
$\mathbf{T}_{6}$~\cite{multibeam2012} 
& 
$\left[
\begin{rsmallmatrix}
1 & 1 & 1 & 1 & 1 & 1 & 1 & 1 \\2 & 1 & 1 & 0 & 0 & -1 & -1 & -2 \\2 & 1 & -1 & -2 & -2 & -1 & 1 & 2 \\1 & 0 & -2 & -1 & 1 & 2 & 0 & -1 \\1 & -1 & -1 & 1 & 1 & -1 & -1 & 1 \\1 & -2 & 0 & 1 & -1 & 0 & 2 & -1 \\1 & -2 & 2 & -1 & -1 & 2 & -2 & 1 \\0 & -1 & 1 & -2 & 2 & -1 & 1 & 0 \\
\end{rsmallmatrix}
\right]$
&
$(3/\gamma_1, 3/\gamma_2)$
&
Yes
\\[25pt]
$\mathbf{T}_{7}$ 
& 
$\left[
\begin{rsmallmatrix}
2 & 2 & 2 & 2 & 2 & 2 & 2 & 2 \\3 & 2 & 1 & 1 & -1 & -1 & -2 & -3 \\2 & 1 & -1 & -2 & -2 & -1 & 1 & 2 \\2 & -1 & -3 & -1 & 1 & 3 & 1 & -2 \\2 & -2 & -2 & 2 & 2 & -2 & -2 & 2 \\1 & -3 & 1 & 2 & -2 & -1 & 3 & -1 \\1 & -2 & 2 & -1 & -1 & 2 & -2 & 1 \\1 & -1 & 2 & -3 & 3 & -2 & 1 & -1 \\
\end{rsmallmatrix}
\right]$
&
$(1/\gamma_6, 3/\gamma_4)$
&
Yes
\\[25pt]
$\tilde{\mathbf{T}}_{1}$
&
See Table~\ref{table-truncation}
&
$(1/\gamma_3,1/\gamma_4)$
&
No
\\
\bottomrule
\end{tabular}
\end{table}

\begin{table}
\centering
\caption{Nearest integer functions}
\label{table-nearest-integer-functions}
\begin{tabular}{ccc}
\toprule
Function
&
Range of $\alpha$ 
&
Transformation matrix
\\
\midrule
\multirow{6}{*}{$\operatorname{round}_\text{HAFZ}(\cdot)$}
&
$[1/\gamma_4, 1/\gamma_5)$
&
$\mathbf{T}_{0}$
\\
&
$[1/\gamma_5, 3/\gamma_0)$
&
$\mathbf{T}_{4}$
\\
&
$(3/\gamma_0, 3/\gamma_1)$
&
$\mathbf{T}_{5}$
\\
&
$(3/\gamma_1, 3/\gamma_2)$
&
$\mathbf{T}_{6}$
\\
&
$[1/\gamma_6, 3/\gamma_4)$
&
$\mathbf{T}_{7}$
\\
&
$[1/\gamma_3, 1/\gamma_4)$
&
$\tilde{\mathbf{T}}_{1}$
\\
\midrule
\multirow{6}{*}{$\operatorname{round}_\text{HTZ}(\cdot)$}
&
$(1/\gamma_4, 1/\gamma_5)$
&
$\mathbf{T}_{0}$
\\
&
$[1/\gamma_5, 3/\gamma_0)$
&
$\mathbf{T}_{4}$
\\
&
$[3/\gamma_0, 3/\gamma_1)$
&
$\mathbf{T}_{5}$
\\
&
$[3/\gamma_1, 3/\gamma_2)$
&
$\mathbf{T}_{6}$
\\
&
$(1/\gamma_6, 3/\gamma_4)$
&
$\mathbf{T}_{7}$
\\
&
$[1/\gamma_3, 1/\gamma_4)$
&
$\tilde{\mathbf{T}}_{1}$
\\
\midrule
\multirow{6}{*}{$\operatorname{round}_\text{EVEN}(\cdot)$}
&
$(1/\gamma_4, 1/\gamma_5)$
&
$\mathbf{T}_{0}$
\\
&
$[1/\gamma_5, 3/\gamma_0)$
&
$\mathbf{T}_{4}$
\\
&
$(3/\gamma_0, 3/\gamma_1)$
&
$\mathbf{T}_{5}$
\\
&
$(3/\gamma_1, 3/\gamma_2)$
&
$\mathbf{T}_{6}$
\\
&
$[1/\gamma_6, 3/\gamma_4)$
&
$\mathbf{T}_{7}$
\\
&
$[1/\gamma_3, 1/\gamma_4)$
&
$\tilde{\mathbf{T}}_{1}$
\\
\midrule
\multirow{6}{*}{$\operatorname{round}_\text{ODD}(\cdot)$}
&
$[1/\gamma_4, 1/\gamma_5)$
&
$\mathbf{T}_{0}$
\\
&
$[1/\gamma_5, 3/\gamma_0)$
&
$\mathbf{T}_{4}$
\\
&
$[3/\gamma_0, 3/\gamma_1)$
&
$\mathbf{T}_{5}$
\\
&
$[3/\gamma_1, 3/\gamma_2)$
&
$\mathbf{T}_{6}$
\\
&
$[1/\gamma_6, 3/\gamma_4)$
&
$\mathbf{T}_{7}$
\\
&
$[1/\gamma_3, 1/\gamma_4)$
&
$\tilde{\mathbf{T}}_{1}$
\\
\bottomrule
\end{tabular}
\end{table}

\begin{table}
\centering
\caption{Orthogonal approximations}
\label{table-diagonal-D}
\begin{tabular}{lc}
\toprule
Approximation $\mathbf{T}$
&
Diagonal elements of $\mathbf{T}\cdot\mathbf{T}^\top$
\\
\midrule
$\mathbf{T}_{0}$
&
$\begin{bmatrix}
8  &   6  &   4  &   6  &   8  &   6  &   4  &   6
\end{bmatrix}$
\\
$\mathbf{T}_{1}$
&
$\begin{bmatrix}
8  &  12  &   4  &  12  &   8  &  12  &   4  &  12
\end{bmatrix}$
\\
$\mathbf{T}_{2}$
&
$\begin{bmatrix}
8  &  12 &   16  &  12 &   8  &  12  &  16 &   12
\end{bmatrix}$
\\
$\mathbf{T}_{3}$
&
$\begin{bmatrix}
32  &  34  &  40  &  34 &   32  &  34  &  40 &   34
\end{bmatrix}$
\\
$\mathbf{T}_{4}$~\cite{cb2011}
&
$\begin{bmatrix}
8  &  6  &  8  &  6 &   8  &  6  &  8 &   6
\end{bmatrix}$
\\
$\mathbf{T}_{5}$
&
$\begin{bmatrix}
8 &   12  &   8 &   12 &    8  &  12 &    8  &  12
\end{bmatrix}$
\\
$\mathbf{T}_{6}$~\cite{multibeam2012}
&
$\begin{bmatrix}
8  &  12  &  20  &  12 &   8  &  12 &  20  &  12
\end{bmatrix}$
\\
$\mathbf{T}_{7}$
&
$\begin{bmatrix}
32  &  30  &  20  & 30  &  32  &  30  &  20  &   30
\end{bmatrix}$
\\
\bottomrule
\end{tabular}
\end{table}

Considering orthogonal matrices,
as shown in~\eqref{eq:approximation},
the
diagonal matrix~$\mathbf{S} = \sqrt{\mathbf{D}}$
is required to
orthonormalize~$\mathbf{T}$.
In Table~\ref{table-diagonal-D},
the required diagonal elements are
listed.
As described in~\cite{bc2012,bas2011,lengwehasatit2004scalable,bas2008,bas2009,cb2011,bayer201216pt},
in the context of JPEG-like 
image compression,
these diagonal matrices represent 
no additional
arithmetic complexity.
This is because they can be absorbed into
the image quantization step~\cite{Wallace1992}.

Regarding the obtained non-orthogonal approximations,
in Table~\ref{table-dfd},
we show the deviation from diagonality values
for the associate 
diagonal matrix~$\mathbf{D}$ as shown in~\eqref{equation-condition}.
The proposed matrices
$\tilde{\mathbf{T}}_{3}$
and
$\tilde{\mathbf{T}}_{4}$
showed
significantly lower
deviation from orthogonality
when compared to the well-known SDCT.

\begin{table}
\centering
\caption{Deviation from diagonality measure}
\label{table-dfd}
\begin{tabular}{lc}
\toprule
Approximation~$\mathbf{T}$
&
$\operatorname{\delta} \left( \mathbf{T} \cdot \mathbf{T}^\top \right)$
\\
\midrule
$\tilde{\mathbf{T}}_{0}$
&
$0.4548$
\\
$\tilde{\mathbf{T}}_{1}$
&
$0.0646$
\\
$\tilde{\mathbf{T}}_{2}$~\cite{haweel2001new}
&
$0.1056$
\\
$\tilde{\mathbf{T}}_{3}$
&
$0.0063$
\\
$\tilde{\mathbf{T}}_{4}$
&
$0.0036$
\\
\bottomrule
\end{tabular}
\end{table}

We explicitly
computed the inverse of the non-orthogonal matrices,
confirming their
low-complexity character.
Indeed,
we have the following inverse matrices:
\begin{align*}
\tilde{\mathbf{T}}_{1}^{-1}
&
=
\left[
\begin{rsmallmatrix}
1 & 1 & 1 & 1 & 1 & 1 & 0 & 1 \\1 & 1 & 0 & -1 & -1 & -1 & -1 & -1 \\1 & 1 & 0 & -1 & -1 & 1 & 1 & 1 \\1 & 1 & -1 & -1 & 1 & 1 & 0 & -1 \\1 & -1 & -1 & 1 & 1 & -1 & 0 & 1 \\1 & -1 & 0 & 1 & -1 & -1 & 1 & -1 \\1 & -1 & 0 & 1 & -1 & 1 & -1 & 1 \\1 & -1 & 1 & -1 & 1 & -1 & 0 & -1 \\
\end{rsmallmatrix}
\right]
\cdot
\left[
\begin{smallmatrix}
1/8 & 0 & 0 & 0 & 0 & 0 & 0 & 0 \\
0 & 1/4 & 0 & 0 & 0 & 0 & 0 & 0 \\
0 & 0 & 1/4 & 0 & 0 & 0 & 0 & 0 \\
0 & 0 & 0 & 1/4 & 0 & 0 & 0 & 0 \\
0 & 0 & 0 & 0 & 1/8 & 0 & 0 & 0 \\
0 & 0 & 0 & 0 & 0 & 1/4 & 0 & 0 \\
0 & 0 & 0 & 0 & 0 & 0 & 1/4 & 0 \\
0 & 0 & 0 & 0 & 0 & 0 & 0 & 1/4 \\
\end{smallmatrix}
\right]
=
\frac{1}{8}
\cdot
\left[
\begin{rsmallmatrix}
1 & 2 & 2 & 2 & 1 & 2 & 0 & 2 \\1 & 2 & 0 & -2 & -1 & -2 & -2 & -2 \\1 & 2 & 0 & -2 & -1 & 2 & 2 & 2 \\1 & 2 & -2 & -2 & 1 & 2 & 0 & -2 \\1 & -2 & -2 & 2 & 1 & -2 & 0 & 2 \\1 & -2 & 0 & 2 & -1 & -2 & 2 & -2 \\1 & -2 & 0 & 2 & -1 & 2 & -2 & 2 \\1 & -2 & 2 & -2 & 1 & -2 & 0 & -2 \\
\end{rsmallmatrix}
\right]
,
\end{align*}

\begin{align*}
\tilde{\mathbf{T}}_{2}^{-1}
&
=
\left[
\begin{rsmallmatrix}
1 & 1 & 1 & 1 & 1 & 0 & 1 & 0 \\1 & 1 & 1 & 0 & -1 & -1 & -1 & 0 \\1 & 0 & -1 & -1 & -1 & 0 & 1 & 1 \\1 & 0 & -1 & 0 & 1 & 1 & -1 & -1 \\1 & 0 & -1 & 0 & 1 & -1 & -1 & 1 \\1 & 0 & -1 & 1 & -1 & 0 & 1 & -1 \\1 & -1 & 1 & 0 & -1 & 1 & -1 & 0 \\1 & -1 & 1 & -1 & 1 & 0 & 1 & 0 \\
\end{rsmallmatrix}
\right]
\cdot
\left[
\begin{smallmatrix}
1/8 & 0 & 0 & 0 & 0 & 0 & 0 & 0 \\
0 & 1/4 & 0 & 0 & 0 & 0 & 0 & 0 \\
0 & 0 & 1/8 & 0 & 0 & 0 & 0 & 0 \\
0 & 0 & 0 & 1/4 & 0 & 0 & 0 & 0 \\
0 & 0 & 0 & 0 & 1/8 & 0 & 0 & 0 \\
0 & 0 & 0 & 0 & 0 & 1/4 & 0 & 0 \\
0 & 0 & 0 & 0 & 0 & 0 & 1/8 & 0 \\
0 & 0 & 0 & 0 & 0 & 0 & 0 & 1/4 \\
\end{smallmatrix}
\right]
=
\frac{1}{8}
\cdot
\left[
\begin{rsmallmatrix}
1 & 2 & 1 & 2 & 1 & 0 & 1 & 0 \\1 & 2 & 1 & 0 & -1 & -2 & -1 & 0 \\1 & 0 & -1 & -2 & -1 & 0 & 1 & 2 \\1 & 0 & -1 & 0 & 1 & 2 & -1 & -2 \\1 & 0 & -1 & 0 & 1 & -2 & -1 & 2 \\1 & 0 & -1 & 2 & -1 & 0 & 1 & -2 \\1 & -2 & 1 & 0 & -1 & 2 & -1 & 0 \\1 & -2 & 1 & -2 & 1 & 0 & 1 & 0 \\
\end{rsmallmatrix}
\right]
,
\end{align*}

\begin{align*}
\tilde{\mathbf{T}}_{3}^{-1}
&
=
\left[
\begin{rsmallmatrix}
1 & 3 & 2 & 3 & 1 & 1 & 1 & 1 \\1 & 3 & 1 & -1 & -1 & -3 & -2 & -1 \\1 & 1 & -1 & -3 & -1 & 1 & 2 & 3 \\1 & 1 & -2 & -1 & 1 & 3 & -1 & -3 \\1 & -1 & -2 & 1 & 1 & -3 & -1 & 3 \\1 & -1 & -1 & 3 & -1 & -1 & 2 & -3 \\1 & -3 & 1 & 1 & -1 & 3 & -2 & 1 \\1 & -3 & 2 & -3 & 1 & -1 & 1 & -1 \\
\end{rsmallmatrix}
\right]
\cdot
\left[
\begin{smallmatrix}
1/8 & 0 & 0 & 0 & 0 & 0 & 0 & 0 \\
0 & 1/28 & 0 & 0 & 0 & 0 & 0 & 0 \\
0 & 0 & 1/20 & 0 & 0 & 0 & 0 & 0 \\
0 & 0 & 0 & 1/28 & 0 & 0 & 0 & 0 \\
0 & 0 & 0 & 0 & 1/8 & 0 & 0 & 0 \\
0 & 0 & 0 & 0 & 0 & 1/28 & 0 & 0 \\
0 & 0 & 0 & 0 & 0 & 0 & 1/20 & 0 \\
0 & 0 & 0 & 0 & 0 & 0 & 0 & 1/28 \\
\end{smallmatrix}
\right]
,
\end{align*}

\begin{align*}
\tilde{\mathbf{T}}_{4}^{-1}
&
=
\left[
\begin{rsmallmatrix}
1 & 3 & 2 & 3 & 1 & 1 & 1 & 1 \\1 & 3 & 1 & -1 & -1 & -3 & -2 & -1 \\1 & 1 & -1 & -3 & -1 & 1 & 2 & 3 \\1 & 1 & -2 & -1 & 1 & 3 & -1 & -3 \\1 & -1 & -2 & 1 & 1 & -3 & -1 & 3 \\1 & -1 & -1 & 3 & -1 & -1 & 2 & -3 \\1 & -3 & 1 & 1 & -1 & 3 & -2 & 1 \\1 & -3 & 2 & -3 & 1 & -1 & 1 & -1 \\
\end{rsmallmatrix}
\right]
\cdot
\left[
\begin{smallmatrix}
1/16 & 0 & 0 & 0 & 0 & 0 & 0 & 0 \\
0 & 1/28 & 0 & 0 & 0 & 0 & 0 & 0 \\
0 & 0 & 1/20 & 0 & 0 & 0 & 0 & 0 \\
0 & 0 & 0 & 1/28 & 0 & 0 & 0 & 0 \\
0 & 0 & 0 & 0 & 1/16 & 0 & 0 & 0 \\
0 & 0 & 0 & 0 & 0 & 1/28 & 0 & 0 \\
0 & 0 & 0 & 0 & 0 & 0 & 1/20 & 0 \\
0 & 0 & 0 & 0 & 0 & 0 & 0 & 1/28 \\
\end{smallmatrix}
\right]
=
\tilde{\mathbf{T}}_{3}^{-1}
\cdot
\left[
\begin{rsmallmatrix}
1/2 & 0 & 0 & 0 & 0 & 0 & 0 & 0 \\
0 & 1 & 0 & 0 & 0 & 0 & 0 & 0 \\
0 & 0 & 1 & 0 & 0 & 0 & 0 & 0 \\
0 & 0 & 0 & 1 & 0 & 0 & 0 & 0 \\
0 & 0 & 0 & 0 & 1/2 & 0 & 0 & 0 \\
0 & 0 & 0 & 0 & 0 & 1 & 0 & 0 \\
0 & 0 & 0 & 0 & 0 & 0 & 1 & 0 \\
0 & 0 & 0 & 0 & 0 & 0 & 0 & 1 \\
\end{rsmallmatrix}
\right]
.
\end{align*}

\subsection{Degenerate Approximations}

By allowing matrix~$\mathbf{D}$ in~\eqref{equation-condition}
to possess null elements in its diagonal,
the resulting approximation matrices may have null rows.
Therefore,
such matrices are non-singular 
and do not furnish good approximations.
We refer to them as degenerate approximations.
However,
the non-null rows of degenerate approximations may be of interest
as simple estimators for their
corresponding particular spectral components.
In Table~\ref{table-degenerate},
we list the obtained degenerate approximations.

\begin{table}
\centering
\caption{Degenerate cases}
\label{table-degenerate}
\begin{tabular}{p{5cm}ccc}
\toprule
Function
&
Range of $\alpha$ 
&
Transformation matrix
&
\\
\midrule
\multirow{10}{*}{$\operatorname{floor}(\cdot)$}
&
$(2,2/\gamma_0)$
&
$\check{\mathbf{T}}_0$
&
$\left[
\begin{rsmallmatrix}
0 & 0 & 0 & 0 & 0 & 0 & 0 & 0 \\0 & 0 & 0 & 0 & -1 & -1 & -1 & -1 \\0 & 0 & -1 & -1 & -1 & -1 & 0 & 0 \\0 & -1 & -1 & -1 & 0 & 0 & 0 & -1 \\0 & -1 & -1 & 0 & 0 & -1 & -1 & 0 \\0 & -1 & 0 & 0 & -1 & -1 & 0 & -1 \\0 & -1 & 0 & -1 & -1 & 0 & -1 & 0 \\0 & -1 & 0 & -1 & 0 & -1 & 0 & -1 \\
\end{rsmallmatrix}
\right]$
\\[25pt]
&
$(2/\gamma_0,2/\gamma_1)$
&
$\check{\mathbf{T}}_1$
&
$\left[
\begin{rsmallmatrix}
0 & 0 & 0 & 0 & 0 & 0 & 0 & 0 \\1 & 0 & 0 & 0 & -1 & -1 & -1 & -2 \\0 & 0 & -1 & -1 & -1 & -1 & 0 & 0 \\0 & -1 & -2 & -1 & 0 & 1 & 0 & -1 \\0 & -1 & -1 & 0 & 0 & -1 & -1 & 0 \\0 & -2 & 0 & 0 & -1 & -1 & 1 & -1 \\0 & -1 & 0 & -1 & -1 & 0 & -1 & 0 \\0 & -1 & 0 & -2 & 1 & -1 & 0 & -1 \\
\end{rsmallmatrix}
\right]$
\\[25pt]
&
$(2/\gamma_1,2/\gamma_2)$
&
$\check{\mathbf{T}}_2$
&
$\left[
\begin{rsmallmatrix}
0 & 0 & 0 & 0 & 0 & 0 & 0 & 0 \\1 & 0 & 0 & 0 & -1 & -1 & -1 & -2 \\1 & 0 & -1 & -2 & -2 & -1 & 0 & 1 \\0 & -1 & -2 & -1 & 0 & 1 & 0 & -1 \\0 & -1 & -1 & 0 & 0 & -1 & -1 & 0 \\0 & -2 & 0 & 0 & -1 & -1 & 1 & -1 \\0 & -2 & 1 & -1 & -1 & 1 & -2 & 0 \\0 & -1 & 0 & -2 & 1 & -1 & 0 & -1 \\
\end{rsmallmatrix}
\right]$
\\[25pt]
\midrule
\multirow{14}{*}{$\operatorname{trunc}(\cdot)$}
&
$2/\gamma_0$
&
$\check{\mathbf{T}}_3$
&
$\left[
\begin{rsmallmatrix}
0 & 0 & 0 & 0 & \phantom{-}0 & \phantom{-}0 & \phantom{-}0 & 0 \\
1 & 0 & 0 & 0 & 0 & 0 & 0 & -1 \\0 & 0 & 0 & 0 & 0 & 0 & 0 & 0 \\0 & 0 & -1 & 0 & 0 & 1 & 0 & 0 \\0 & 0 & 0 & 0 & 0 & 0 & 0 & 0 \\0 & -1 & 0 & 0 & 0 & 0 & 0 & 0 \\0 & 0 & 0 & 0 & 0 & 0 & 0 & 0 \\0 & 0 & 0 & -1 & 1 & 0 & 0 & 0 \\
\end{rsmallmatrix}
\right]$
\\[25pt]
&
$(2/\gamma_0, 2/\gamma_1)$
&
$\check{\mathbf{T}}_4$
&
$\left[
\begin{rsmallmatrix}
0 & 0 & 0 & 0 & \phantom{-}0 & \phantom{-}0 & \phantom{-}0 & 0 \\
1 & 0 & 0 & 0 & 0 & 0 & 0 & -1 \\
0 & 0 & 0 & 0 & 0 & 0 & 0 & 0 \\
0 & 0 & -1 & 0 & 0 & 1 & 0 & 0 \\
0 & 0 & 0 & 0 & 0 & 0 & 0 & 0 \\
0 & -1 & 0 & 0 & 0 & 0 & 1 & 0 \\
0 & 0 & 0 & 0 & 0 & 0 & 0 & 0 \\
0 & 0 & 0 & -1 & 1 & 0 & 0 & 0 \\
\end{rsmallmatrix}
\right]$
\\[25pt]
&
$(2/\gamma_1, 2/\gamma_2)$
&
$\check{\mathbf{T}}_5$
&
$\left[
\begin{rsmallmatrix}
0 & 0 & 0 & 0 & 0 & \phantom{-}0 & 0 & 0 \\1 & 0 & 0 & 0 & 0 & 0 & 0 & -1 \\1 & 0 & 0 & -1 & -1 & 0 & 0 & 1 \\0 & 0 & -1 & 0 & 0 & 1 & 0 & 0 \\0 & 0 & 0 & 0 & 0 & 0 & 0 & 0 \\0 & -1 & 0 & 0 & 0 & 0 & 1 & 0 \\0 & -1 & 1 & 0 & 0 & 1 & -1 & 0 \\0 & 0 & 0 & -1 & 1 & 0 & 0 & 0 \\
\end{rsmallmatrix}
\right]$
\\
&
$(2/\gamma_2, 2/\gamma_3)$
&
$\check{\mathbf{T}}_6$ 
&
$\left[
\begin{rsmallmatrix}
1 & 1 & 1 & 1 & 1 & 1 & 1 & 1 \\1 & 1 & 0 & 0 & 0 & 0 & -1 & -1 \\1 & 0 & 0 & -1 & -1 & 0 & 0 & 1 \\1 & 0 & -1 & 0 & 0 & 1 & 0 & -1 \\1 & -1 & -1 & 1 & 1 & -1 & -1 & 1 \\0 & -1 & 0 & 1 & -1 & 0 & 1 & 0 \\0 & -1 & 1 & 0 & 0 & 1 & -1 & 0 \\0 & 0 & 1 & -1 & 1 & -1 & 0 & 0 \\
\end{rsmallmatrix}
\right]$
\\[25pt]
\midrule
\multirow{1}{*}{\parbox{4cm}{$\operatorname{round}_\text{HU}(\cdot)$, $\operatorname{round}_\text{HD}(\cdot)$}}
&
$1/\gamma_0$
&
$\check{\mathbf{T}}_7$
&
$\left[
\begin{rsmallmatrix}
0 & \phantom{-}0 & \phantom{-}0 & \phantom{-}0 & \phantom{-}0 & \phantom{-}0 & \phantom{-}0 & \phantom{-}0 \\
1 & 0 & 0 & 0 & 0 & 0 & 0 & 0 \\0 & 0 & 0 & 0 & 0 & 0 & 0 & 0 \\0 & 0 & 0 & 0 & 0 & 1 & 0 & 0 \\0 & 0 & 0 & 0 & 0 & 0 & 0 & 0 \\0 & 0 & 0 & 0 & 0 & 0 & 1 & 0 \\0 & 0 & 0 & 0 & 0 & 0 & 0 & 0 \\0 & 0 & 0 & 0 & 1 & 0 & 0 & 0 \\
\end{rsmallmatrix}
\right]$
\\
\midrule
\multirow{1}{*}{\parbox{4.5cm}{\raggedright $\operatorname{round}_\text{HU}(\cdot)$, $\operatorname{round}_\text{HD}(\cdot)$, $\operatorname{round}_\text{EVEN}(\cdot)$, $\operatorname{round}_\text{ODD}(\cdot)$}}
&
$(1/\gamma_0, 1/\gamma_1)$
&
$\check{\mathbf{T}}_4$
&
As above
\\
&
$(1/\gamma_1, 1/\gamma_2)$
&
$\check{\mathbf{T}}_5$
&
As above
\\
\midrule
\multirow{2}{*}{\parbox{4cm}{$\operatorname{round}_\text{HAFZ}(\cdot)$, $\operatorname{round}_\text{ODD}(\cdot)$}}
&
$[1/\gamma_0, 1/\gamma_1)$
&
$\check{\mathbf{T}}_4$
&
As above
\\
&
$[1/\gamma_1, 1/\gamma_2)$
&
$\check{\mathbf{T}}_5$
&
As above
\\
\bottomrule
\end{tabular}
\end{table}

\section{Fast Algorithm}
\label{sec:Fast_Algorithm}

\begin{figure}

\centering
\subfigure[Full diagram]{\scalebox{1}{\input{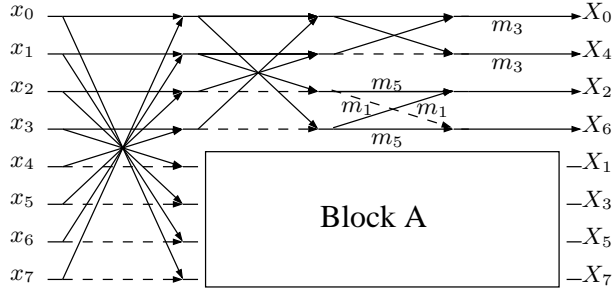}}} \\
\subfigure[Block A]{\scalebox{1}{\input{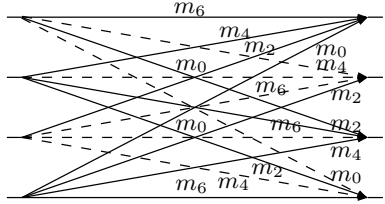}}}
\caption{
General signal flow graph for proposed transformations.
Input data $x_n$, $n=0,1,\ldots,7$, 
relates to output $X_k$, $k=0,1,\ldots,7$, 
according to $\mathbf{X} = \mathbf{T} \cdot \mathbf{x}$.
Dashed arrows represent multiplication by $-1$.}
\label{fig:fast}
\end{figure}

Considering usual decimation-based
techniques
and matrix factorization~\cite{blahut},
fast algorithms for the obtained transformations
could be derived.
All discussed matrices share the
same factorization structure
described below:
\begin{align*}
\mathbf{T}
=%
\mathbf{P}
\cdot
\mathbf{K} 
\cdot 
\mathbf{B}_1 
\cdot 
\mathbf{B}_2 
\cdot \mathbf{B}_3
,
\end{align*}
where
$\mathbf{P}$ is a permutation matrix,
$\mathbf{K}$ is a multiplicative matrix,
and
$\mathbf{B}_1$,
$\mathbf{B}_2$,
and
$\mathbf{B}_3$
are additive matrices.
These matrices are given by:
\begin{align*}
\mathbf{P} &=
\begin{bmatrix}
\begin{smallmatrix}
1 &\phantom{-}0 &\phantom{-}0 & \phantom{-}0 & \phantom{-}0 &\phantom{-}0 &\phantom{-}0 &\phantom{-}0 \\
0 &\phantom{-}0 &\phantom{-}0 & \phantom{-}0 & -1 &\phantom{-}0 &\phantom{-}0 &\phantom{-}0 \\
0 &\phantom{-}0 &\phantom{-}1 & \phantom{-}0 & \phantom{-}0 &\phantom{-}0 &\phantom{-}0 &\phantom{-}0 \\
0 &\phantom{-}0 &\phantom{-}0 & \phantom{-}0 & \phantom{-}0 &-1 &\phantom{-}0 &\phantom{-}0 \\
0 &\phantom{-}1 &\phantom{-}0 & \phantom{-}0 & \phantom{-}0 &\phantom{-}0 &\phantom{-}0 &\phantom{-}0 \\
0 &\phantom{-}0 &\phantom{-}0 & \phantom{-}0 & \phantom{-}0 &\phantom{-}0 &\phantom{-}0 &-1 \\
0 &\phantom{-}0 &\phantom{-}0 & \phantom{-}1 & \phantom{-}0 &\phantom{-}0 &\phantom{-}0 &\phantom{-}0 \\
0 &\phantom{-}0 &\phantom{-}0 & \phantom{-}0 & \phantom{-}0 &\phantom{-}0 &\phantom{-}1 &\phantom{-}0
\end{smallmatrix}
\end{bmatrix}
,
\qquad
\mathbf{K} 
=
\begin{bmatrix}
\begin{smallmatrix}
{m}_3 &\phantom{-}0 &\phantom{-}0 & \phantom{-}0 & \phantom{-}0 &\phantom{-}0 &\phantom{-}0 &\phantom{-}0 \\
0 & {m}_3 &\phantom{-}0 & \phantom{-}0 & \phantom{-}0 &\phantom{-}0 &\phantom{-}0 &\phantom{-}0 \\
0 &\phantom{-}0 & \phantom{-}{m}_5 & \phantom{-}{m}_1 & \phantom{-}0 &\phantom{-}0 &\phantom{-}0 &\phantom{-}0 \\
0 &\phantom{-}0 & -{m}_1 & \phantom{-}{m}_5 & \phantom{-}0 &\phantom{-}0 &\phantom{-}0 &\phantom{-}0 \\
0 &\phantom{-}0 &\phantom{-}0 & \phantom{-}0 & \phantom{-}{m}_4 & -{m}_6 & \phantom{-}{m}_2 & \phantom{-}{m}_0 \\
0 &\phantom{-}0 &\phantom{-}0 & \phantom{-}0 & -{m}_0 & \phantom{-}{m}_4 & -{m}_6 & \phantom{-}{m}_2 \\
0 &\phantom{-}0 &\phantom{-}0 & \phantom{-}0 & -{m}_2 & -{m}_0 & \phantom{-}{m}_4 & -{m}_6 \\
0 &\phantom{-}0 &\phantom{-}0 & \phantom{-}0 & \phantom{-}{m}_6 & -{m}_2 & -{m}_0 & \phantom{-}{m}_4 \\
\end{smallmatrix}
\end{bmatrix}
,
\end{align*}

\begin{align*}
\mathbf{B}_1 &=
\begin{bmatrix}
\begin{smallmatrix}
1 &\phantom{-}1 &\phantom{-}0 & \phantom{-}0 & \phantom{-}0 &\phantom{-}0 &\phantom{-}0 &\phantom{-}0 \\
1 & -1 &\phantom{-}0 & \phantom{-}0 & \phantom{-}0 &\phantom{-}0 &\phantom{-}0 &\phantom{-}0 \\
0 &\phantom{-}0 &\phantom{-}0 & \phantom{-}1 & \phantom{-}0 &\phantom{-}0 &\phantom{-}0 &\phantom{-}0 \\
0 &\phantom{-}0 &\phantom{-}1 & \phantom{-}0 & \phantom{-}0 &\phantom{-}0 &\phantom{-}0 &\phantom{-}0 \\
0 &\phantom{-}0 &\phantom{-}0 & \phantom{-}0 & \phantom{-}0	&\phantom{-}0 & -1 &\phantom{-}0 \\
0 &\phantom{-}0 &\phantom{-}0 & \phantom{-}0 & \phantom{-}0 &\phantom{-}0 &\phantom{-}0 &\phantom{-}1 \\
0 &\phantom{-}0 &\phantom{-}0 & \phantom{-}0 & \phantom{-}0 & -1 & \phantom{-}0 &\phantom{-}0 \\
0 &\phantom{-}0 &\phantom{-}0 & \phantom{-}0 & -1 &\phantom{-}0 &\phantom{-}0 & \phantom{-}0 
\end{smallmatrix}
\end{bmatrix}
,
\qquad
\mathbf{B}_2 
=
\begin{bmatrix}
\begin{smallmatrix}
1 &\phantom{-}0 &\phantom{-}0 & \phantom{-}1 & \phantom{-}0 &\phantom{-}0 &\phantom{-}0 &\phantom{-}0 \\
0 &\phantom{-}1 &\phantom{-}1 & \phantom{-}0 & \phantom{-}0 &\phantom{-}0 &\phantom{-}0 &\phantom{-}0 \\
1 &\phantom{-}0 &\phantom{-}0 & -1           & \phantom{-}0 &\phantom{-}0 &\phantom{-}0 &\phantom{-}0 \\
0 &\phantom{-}1 & -1          & \phantom{-}0 & \phantom{-}0 &\phantom{-}0 &\phantom{-}0 &\phantom{-}0 \\
0 &\phantom{-}0 &\phantom{-}0 & \phantom{-}0 & \phantom{-}1           &\phantom{-}0 &\phantom{-}0 &\phantom{-}0 \\
0 &\phantom{-}0 &\phantom{-}0 & \phantom{-}0 & \phantom{-}0 &\phantom{-}1          &\phantom{-}0 &\phantom{-}0 \\
0 &\phantom{-}0 &\phantom{-}0 & \phantom{-}0 & \phantom{-}0 &\phantom{-}0 &\phantom{-}1          &\phantom{-}0 \\
0 &\phantom{-}0 &\phantom{-}0 & \phantom{-}0 & \phantom{-}0 &\phantom{-}0 &\phantom{-}0 &\phantom{-}1
\end{smallmatrix}
\end{bmatrix}
,
\end{align*}
and
\begin{align*}
\mathbf{B}_3 &=
\begin{bmatrix}
\begin{smallmatrix}
1 &\phantom{-}0 &\phantom{-}0 & \phantom{-}0 & \phantom{-}0 &\phantom{-}0 &\phantom{-}0 &\phantom{-}1 \\
0 &\phantom{-}1 &\phantom{-}0 & \phantom{-}0 & \phantom{-}0 &\phantom{-}0 &\phantom{-}1 &\phantom{-}0 \\
0 &\phantom{-}0 &\phantom{-}1 & \phantom{-}0 & \phantom{-}0 &\phantom{-}1 &\phantom{-}0 &\phantom{-}0 \\
0 &\phantom{-}0 &\phantom{-}0 & \phantom{-}1 & \phantom{-}1 &\phantom{-}0 &\phantom{-}0 &\phantom{-}0 \\
1 &\phantom{-}0 &\phantom{-}0 & \phantom{-}0 & \phantom{-}0 &\phantom{-}0 &\phantom{-}0 &-1 \\
0 &\phantom{-}1 &\phantom{-}0 & \phantom{-}0 & \phantom{-}0 &\phantom{-}0 &-1 &\phantom{-}0 \\
0 &\phantom{-}0 &\phantom{-}1 & \phantom{-}0 & \phantom{-}0 &-1 &\phantom{-}0 &\phantom{-}0 \\
0 &\phantom{-}0 &\phantom{-}0 & \phantom{-}1 & -1 &\phantom{-}0 &\phantom{-}0 &\phantom{-}0
\end{smallmatrix}
\end{bmatrix}
,
\end{align*}
where constants
$m_i$,
$i=0,1,\ldots,6$,
depend on the 
particular choice of
transformation matrix~$\mathbf{T}$.
In
Table~\ref{table-constants},
constants $m_i$,
$i=0,1,\ldots,6$,
are listed 
for each of the discussed transformations.
As a consequence,
all transformations
also share the same 
fast algorithm
and signal flow structure,
as presented in Fig.~\ref{fig:fast}.
For each transform,
we could assess the arithmetic complexity,
as measured by
multiplication,
addition,
and bit-shifting operation
counts.
A multiplication by 3 was counted as one addition and one bit-shift operation.
Results
are shown
in Table~\ref{table-complexity}.
All proposed algorithms are multiplierless;
requiring only additions and bit-shifts operations.

As shown in~\eqref{equation-equivalence},
transforms $\tilde{\mathbf{T}}_3$ and $\tilde{\mathbf{T}}_4$
lead to same approximations.
Since~$\tilde{\mathbf{T}}_4$
requires more arithmetic operations than
$\tilde{\mathbf{T}}_3$,
we 
do not consider~$\tilde{\mathbf{T}}_4$
for further analysis.

\begin{table}
\centering
\caption{Constants required for the fast algorithm}
\label{table-constants}
\begin{tabular}{lcccccccc}
\toprule
Approximation
&
$m_0$ & $m_1$ & $m_2$ & $m_3$ & 
$m_4$ & $m_5$ & $m_6$
\\
\midrule
$\mathbf{T}_{0}$
& 
1 & 1 & 1 & 1 & 1 & 0 & 0\\
$\mathbf{T}_{1}$
& 
2 & 0 & 1 & 1 & 1 & 1 & 0\\
$\mathbf{T}_{2}$
& 
2 & 2 & 1 & 1 & 1 & 0 & 0\\
$\mathbf{T}_{3}$
& 
3 & 3 & 2 & 2 & 2 & 1 & 0\\
$\mathbf{T}_{4}$~\cite{cb2011}
& 
1 & 1 & 1 & 1 & 1 & 1 & 0\\
$\mathbf{T}_{5}$
& 
2 & 1 & 1 & 1 & 1 & 1 & 0\\
$\mathbf{T}_{6}$~\cite{multibeam2012}
& 
2 & 2 & 1 & 1 & 1 & 1 & 0\\
$\mathbf{T}_{7}$
& 
3 & 2 & 2 & 2 & 1 & 1 & 1\\
$\tilde{\mathbf{T}}_{1}$
& 
1 & 1 & 1 & 1 & 0 & 0 & 0\\
$\tilde{\mathbf{T}}_{2}$~\cite{haweel2001new}
& 
1 & 1 & 1 & 1 & 1 & 1 & 1\\
$\tilde{\mathbf{T}}_{3}$
& 
2 & 2 & 2 & 1 & 1 & 1 & 1\\
$\tilde{\mathbf{T}}_{4}$
& 
2 & 2 & 2 & 2 & 1 & 1 & 1\\
\bottomrule
\end{tabular}
\end{table}

\begin{table}
\centering
\caption{Arithmetic complexity of the obtained approximations}
\label{table-complexity}
\begin{tabular}{lccc}
\toprule
Approximation & Multiplications & Additions & Bit-shifts \\
\midrule
$\mathbf{T}_0$ & 0 & 22 & 0 \\
$\mathbf{T}_1$ & 0 & 22 & 4 \\
$\mathbf{T}_2$ & 0 & 22 & 6 \\
$\mathbf{T}_3$ & 0 & 30 & 16 \\
$\mathbf{T}_4$~\cite{cb2011} & 0 & 24 & 0 \\
$\mathbf{T}_5$ & 0 & 24 & 4 \\
$\mathbf{T}_6$~\cite{multibeam2012} & 0 & 24 & 6 \\
$\mathbf{T}_7$ & 0 & 32 & 12 \\
$\tilde{\mathbf{T}}_1$ & 0 & 18 & 0 \\
$\tilde{\mathbf{T}}_2$~\cite{haweel2001new} & 0 & 28 & 0 \\
$\tilde{\mathbf{T}}_3$ & 0 & 28 & 10 \\
$\tilde{\mathbf{T}}_4$ & 0 & 28 & 12 \\
\bottomrule
\end{tabular}
\end{table}

\section{Image Compression}
\label{sec:Assessment}

\subsection{JPEG-like Compression}

Discussed transformations
were considered as tools for 
JPEG-like image compression.
Adopting
the
computational experiment described in~\cite{cb2011,bc2012,bayer201216pt},
we
employed 45 $512\times512$ 8-bit images
obtained from a public image bank~\cite{uscsipi}.
All images were subdivided in 8$\times$8 blocks and were submitted 
to 
a
\mbox{2-D} transformation
similar to~\eqref{equation-2d-transformation},
where the
exact DCT matrix
is replaced
with
a selected DCT approximation. 
The resulting 64 coefficients in transform domain
were ordered in the standard zigzag sequence~\cite{Wallace1992}.
Only the $r$ initial coefficients 
in each block 
were retained;
being the remaining coefficients discarded.
We adopted $1\leq r\leq45$.

Subsequently,
the inverse \mbox{2-D} transform was applied
and the compressed images were obtained. 
Original and compressed images
were then evaluated for image degradation.
As
quality assessment measures,
we considered
the peak signal-to-noise ratio (PSNR)~\cite{Huynh-Thu2008} 
and
the structural similarity index (SSIM)~\cite{Wang2004}.
For each value of $r$,
average image quality measures based on the 45 images
were considered.
As opposed to analyzing particular images as in~\cite{bas2008, bouguezel2008multiplication, bas2009, bas2010, bas2011},
by taking average measurements,
the suggested approach is less prone to variance effects
and fortuitous data.
Therefore,
such methodology is more robust~\cite{kay1993,cb2011}.

\subsection{Results and Discussion}

\begin{figure*}
\centering

 \subfigure[]
 {\includegraphics[width=0.48\linewidth]{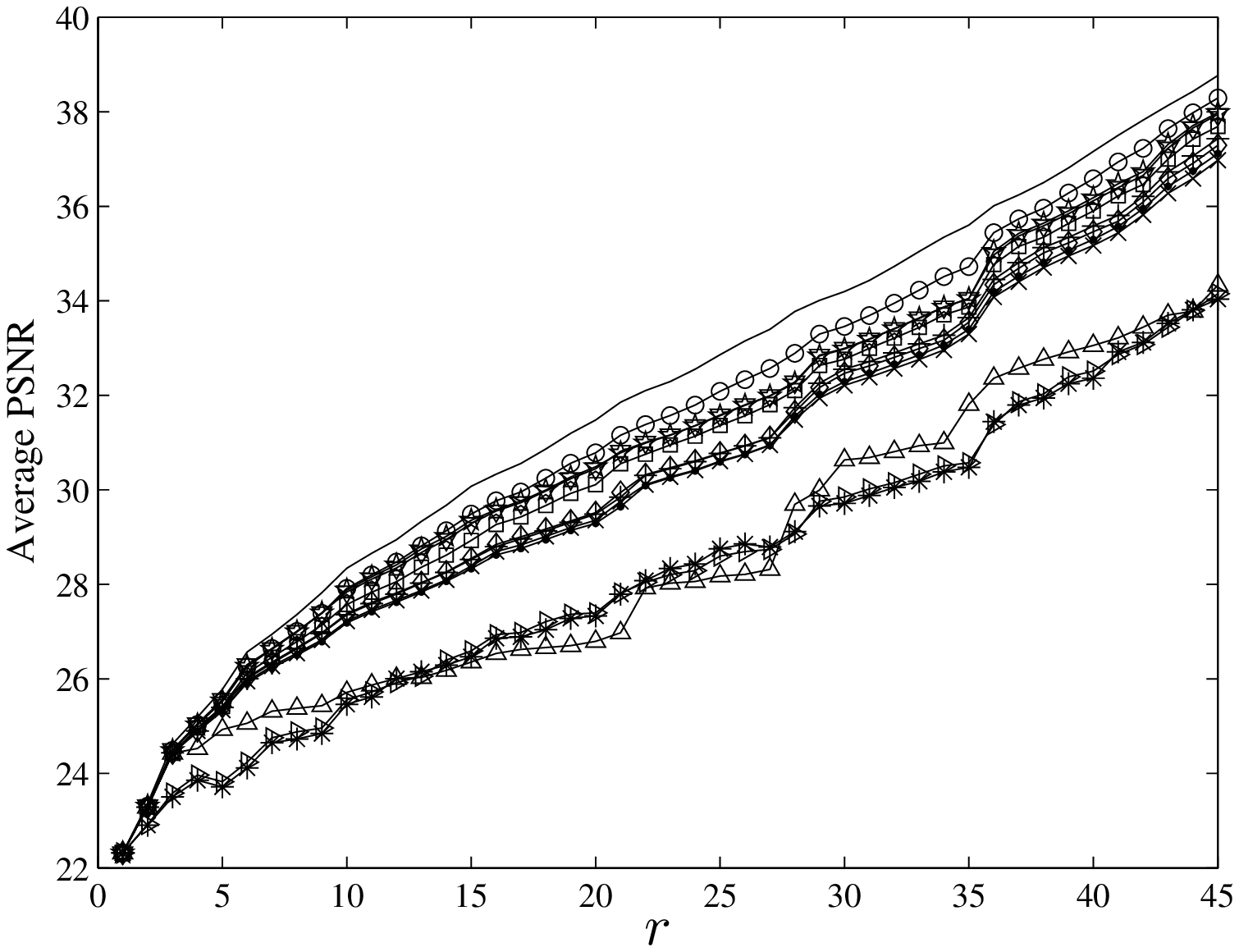}\label{psnr}}
 \subfigure[]
 {\psfrag{aaaa}{\scriptsize DCT}
 \psfrag{bbbb}{\scriptsize $\mathbf{T}_0$}
 \psfrag{cccc}{\scriptsize $\mathbf{T}_1$}
 \psfrag{dddd}{\scriptsize $\mathbf{T}_2$}
 \psfrag{eeee}{\scriptsize $\mathbf{T}_3$}
 \psfrag{ffff}{\scriptsize $\mathbf{T}_4$~\cite{cb2011}}
 \psfrag{gggg}{\scriptsize $\mathbf{T}_5$}
 \psfrag{hhhh}{\scriptsize $\mathbf{T}_6$~\cite{multibeam2012}}
 \psfrag{iiii}{\scriptsize $\mathbf{T}_7$}
 \psfrag{jjjj}{\scriptsize $\tilde{\mathbf{T}}_1$}
 \psfrag{kkkk}{\scriptsize $\tilde{\mathbf{T}}_2$~\cite{haweel2001new}}
 \psfrag{llll}{\scriptsize $\tilde{\mathbf{T}}_3$}
 \includegraphics[width=0.48\linewidth]{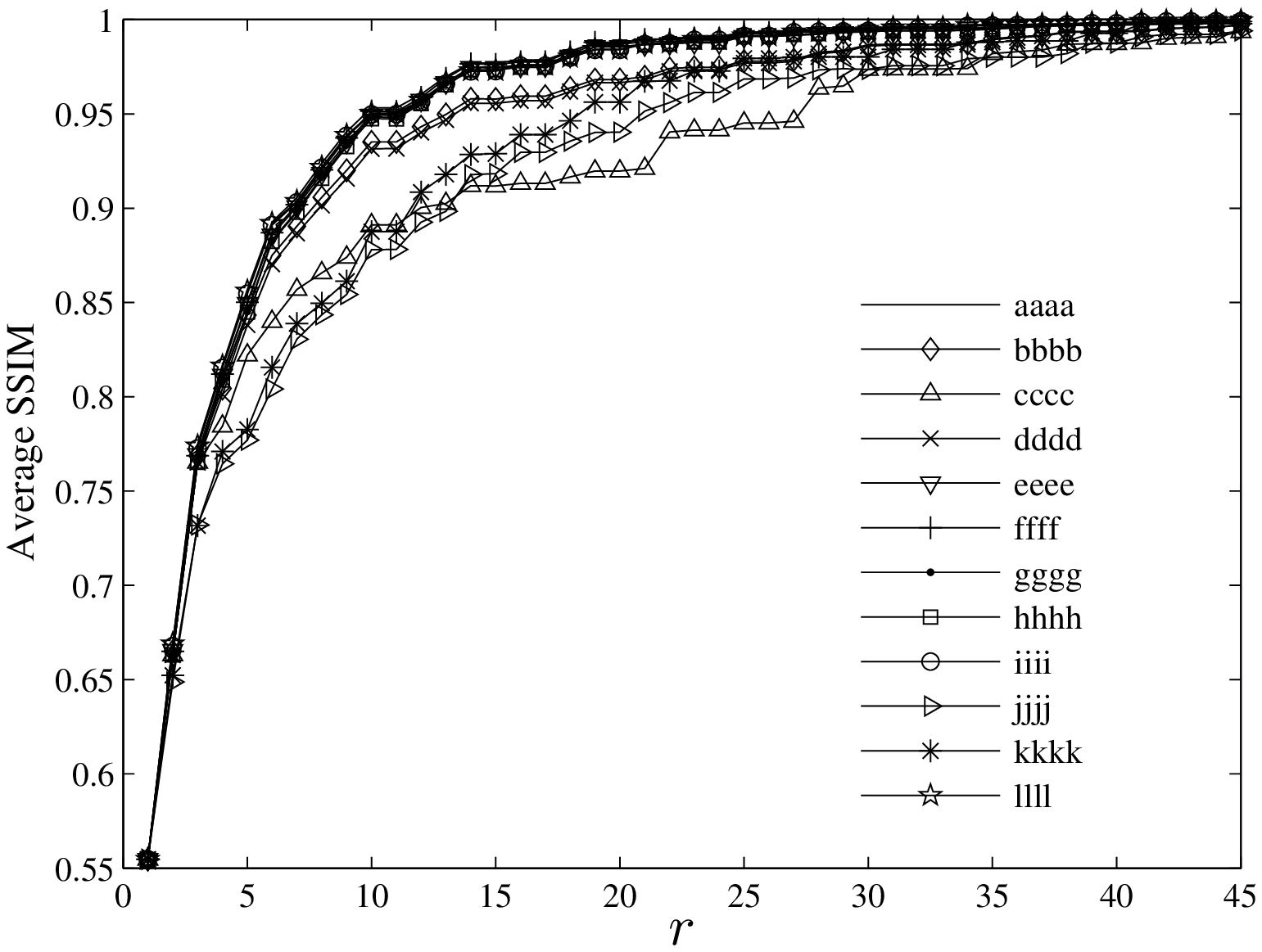}\label{ssim}} 
 \subfigure[]
 {\includegraphics[width=0.48\linewidth]{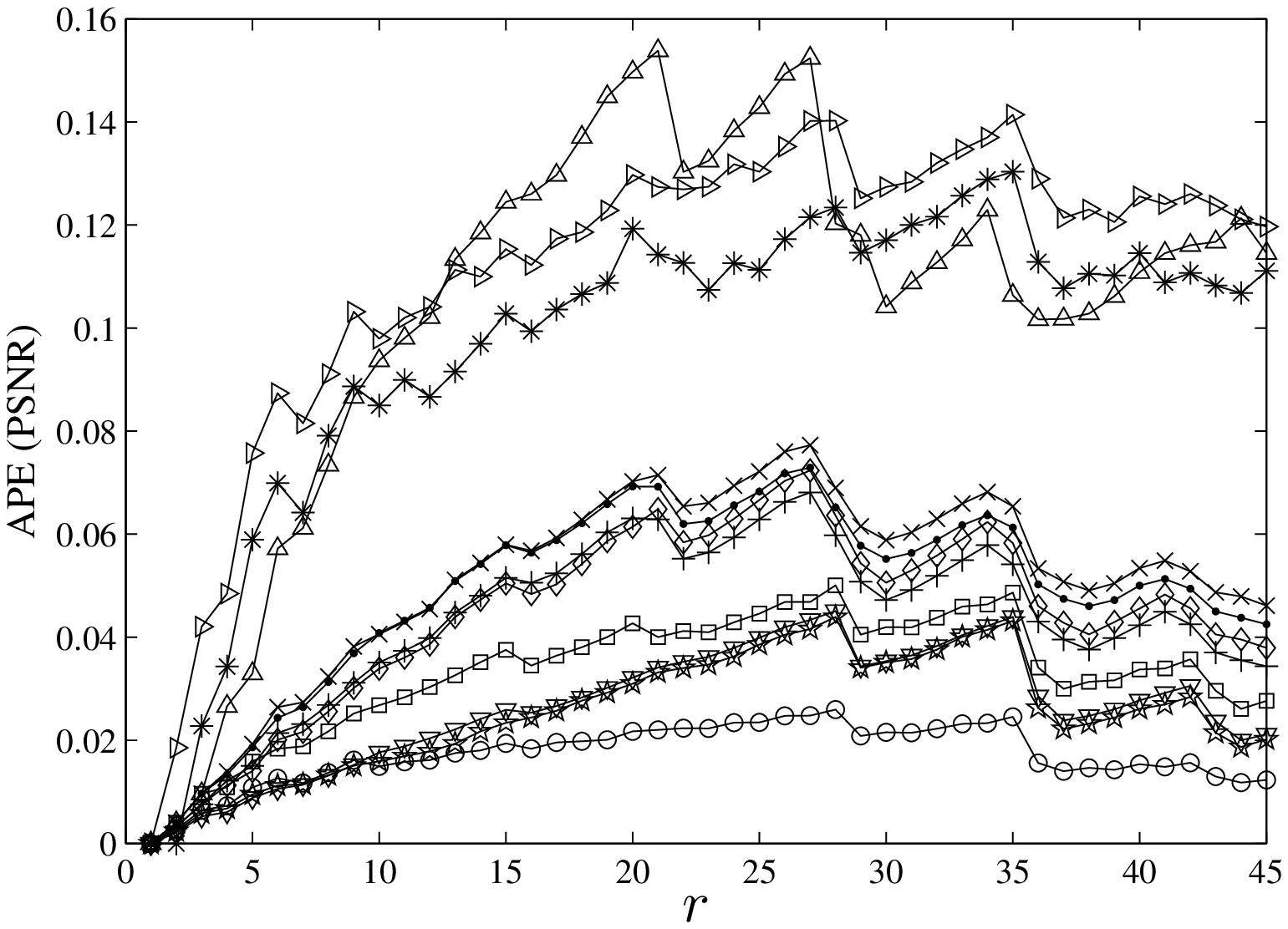}\label{apepsnr}}
 \subfigure[]
 {\includegraphics[width=0.48\linewidth]{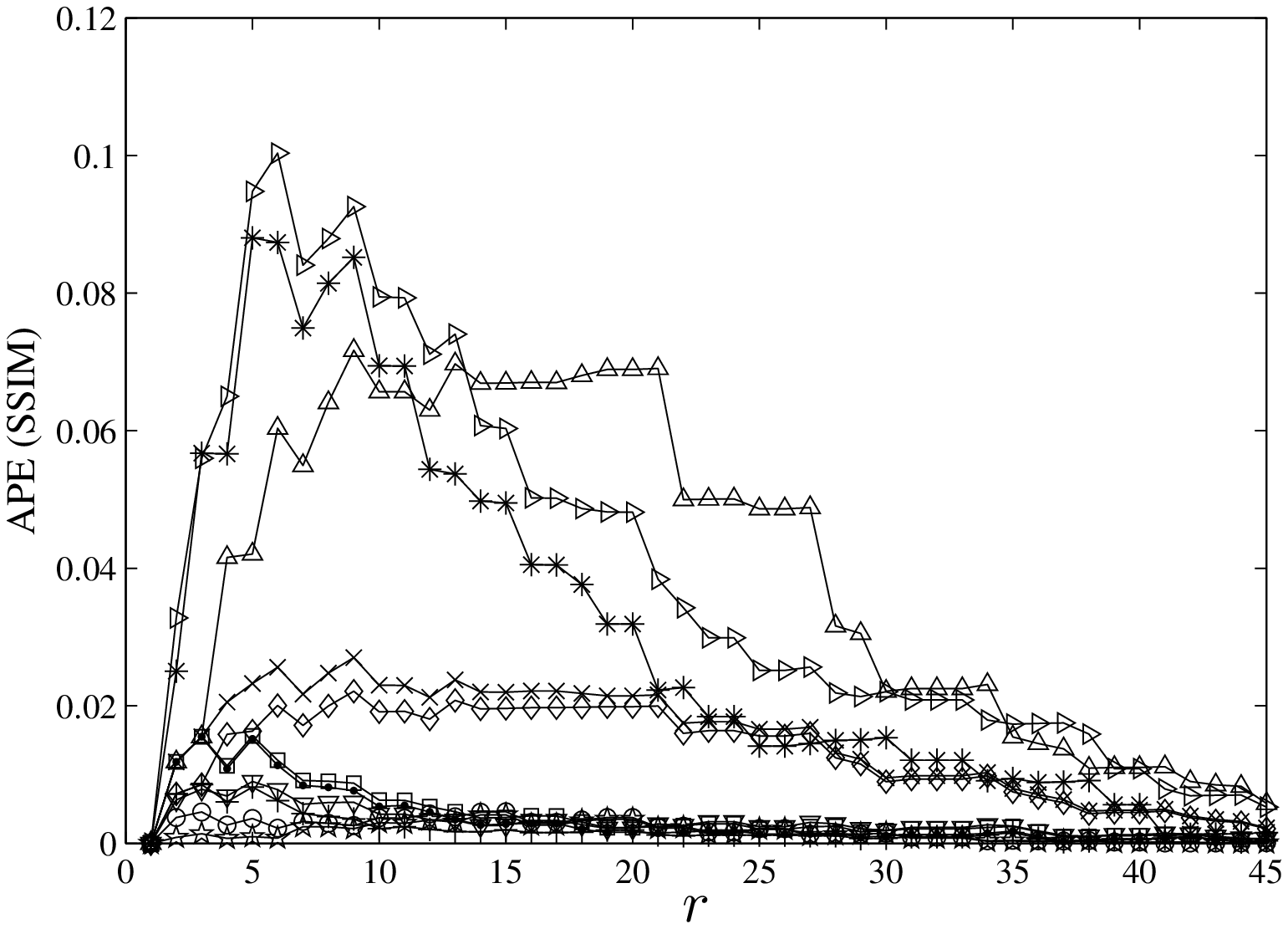}\label{apessim}}
\caption{
Quality measures of 
the considered approximations
for several values of $r$ according to the following
figures of merit:
\subref{psnr}~Average PSNR, 
\subref{ssim}~Average SSIM, 
\subref{apepsnr}~Average PSNR absolute percentage error relative to the DCT, 
and 
\subref{apessim}~Average SSIM absolute percentage error relative to the DCT.}
 \label{fig:measures}
\end{figure*}

Fig.~\ref{fig:measures}(a)-(b)
show
the obtained plots based on the selected quality assessment measures.
In order to enhance visualization
of the results, 
we considered the absolute percentage error~(APE) 
relative to the DCT 
as shown in Fig.~\ref{fig:measures}(c)-(d).

Approximation~$\mathbf{T}_7$
outperformed all other approximations.
This is partially expected
because,
by using all possible elements in $\mathcal{C}$,
it may potentially better approximate the actual
DCT vector basis.
Nevertheless,
it should be noticed
that~$\mathbf{T}_7$
also possesses
the highest computational cost among the examined transformations.
On the other hand,
approximation~$\tilde{\mathbf{T}}_1$ has the lowest computational complexity, requiring only 18 additions. 
The orthogonal approximation $\mathbf{T}_3$ 
showed comparable performance to the 
non-orthogonal approximation $\tilde{\mathbf{T}}_3$.
However,
the non-orthogonal approximation
is less computationally expensive requiring 28 additions and 10 bit-shifts; whereas~$\mathbf{T}_3$ requires 30 additions and 16 bit-shifting operations. 
Comparing the approximations
with 22 additions, 
we have that $\mathbf{T}_0$ 
could outperform
approximations $\mathbf{T}_1$ and $\mathbf{T}_2$ 
in terms of PSNR and SSIM measures
for all considered values of $r$.
In terms of the approximations with 24 additions, 
we have that $\mathbf{T}_4$ 
showed better behavior
than approximations $\mathbf{T}_5$ and $\mathbf{T}_6$
according to both measures.
Moreover,
$\mathbf{T}_4$ requires no bit-shifting operations. 
Focusing on the non-orthogonal transforms, 
$\tilde{\mathbf{T}}_3$ presented
the best performance in terms of PSNR and SSIM measures.
However,
$\tilde{\mathbf{T}}_1$ and SDCT
showed lower computational complexity.

\begin{figure*}

\centering
\subfigure[$\mathbf{T}_0$ (PSNR=27.862, SSIM=0.955)]
{\includegraphics[width=0.2\linewidth]{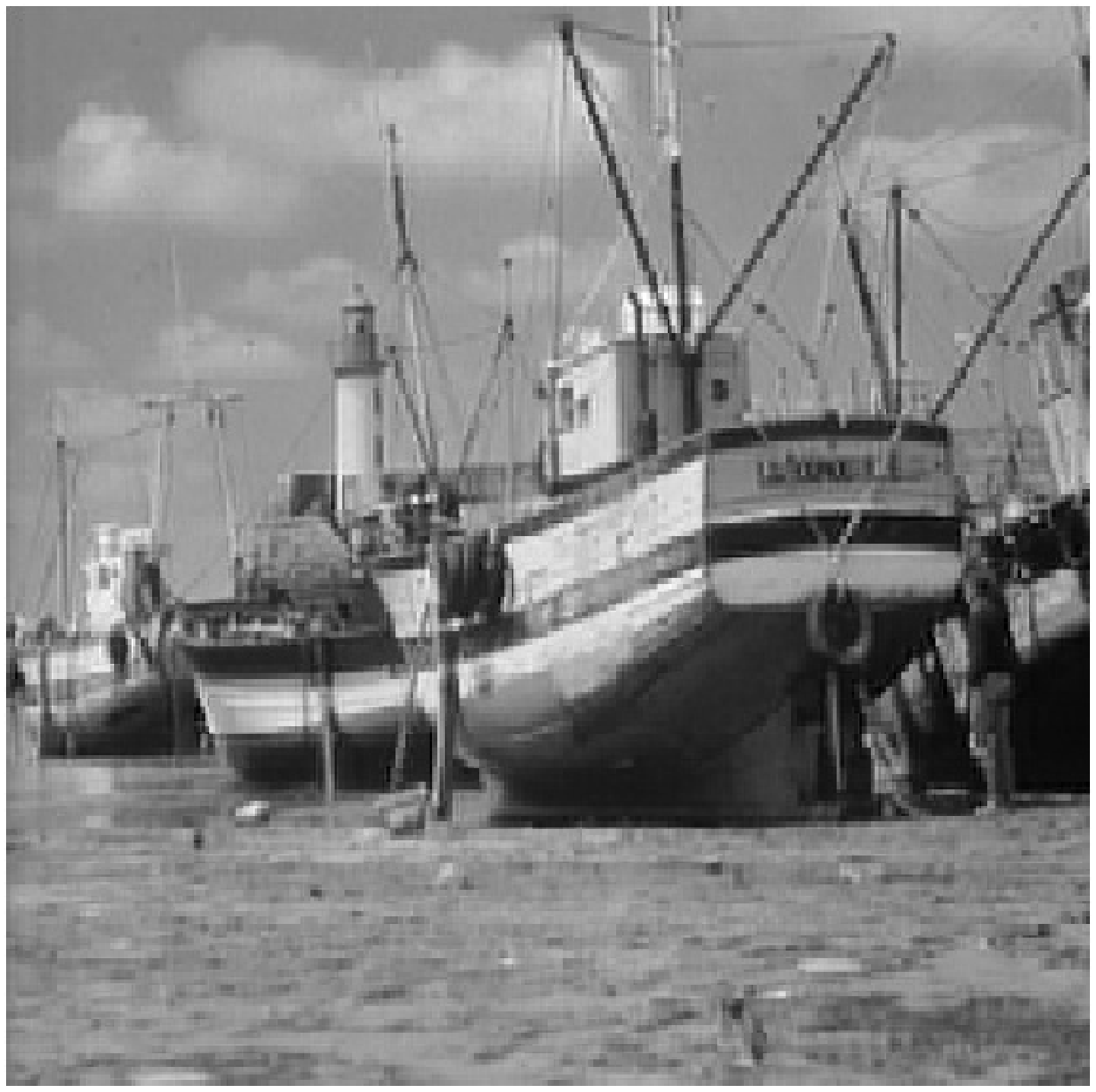} \label{boat_t0}}
\subfigure[$\mathbf{T}_4$~\cite{cb2011} (PSNR=27.870, SSIM=0.968)]
{\includegraphics[width=0.2\linewidth]{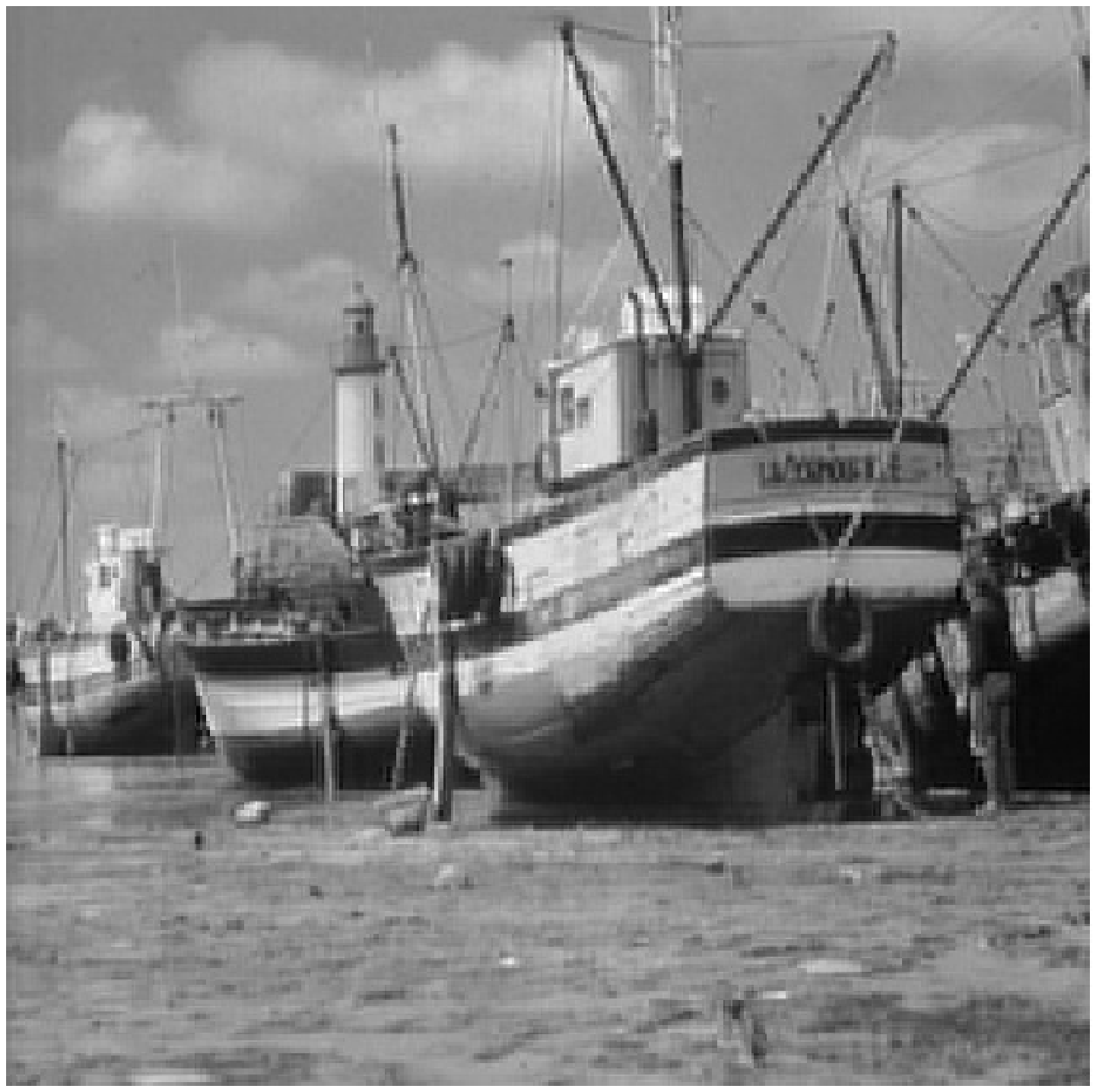} \label{boat_t4}}
\subfigure[$\tilde{\mathbf{T}}_1$ (PSNR=25.805, SSIM=0.903)]
{\includegraphics[width=0.2\linewidth]{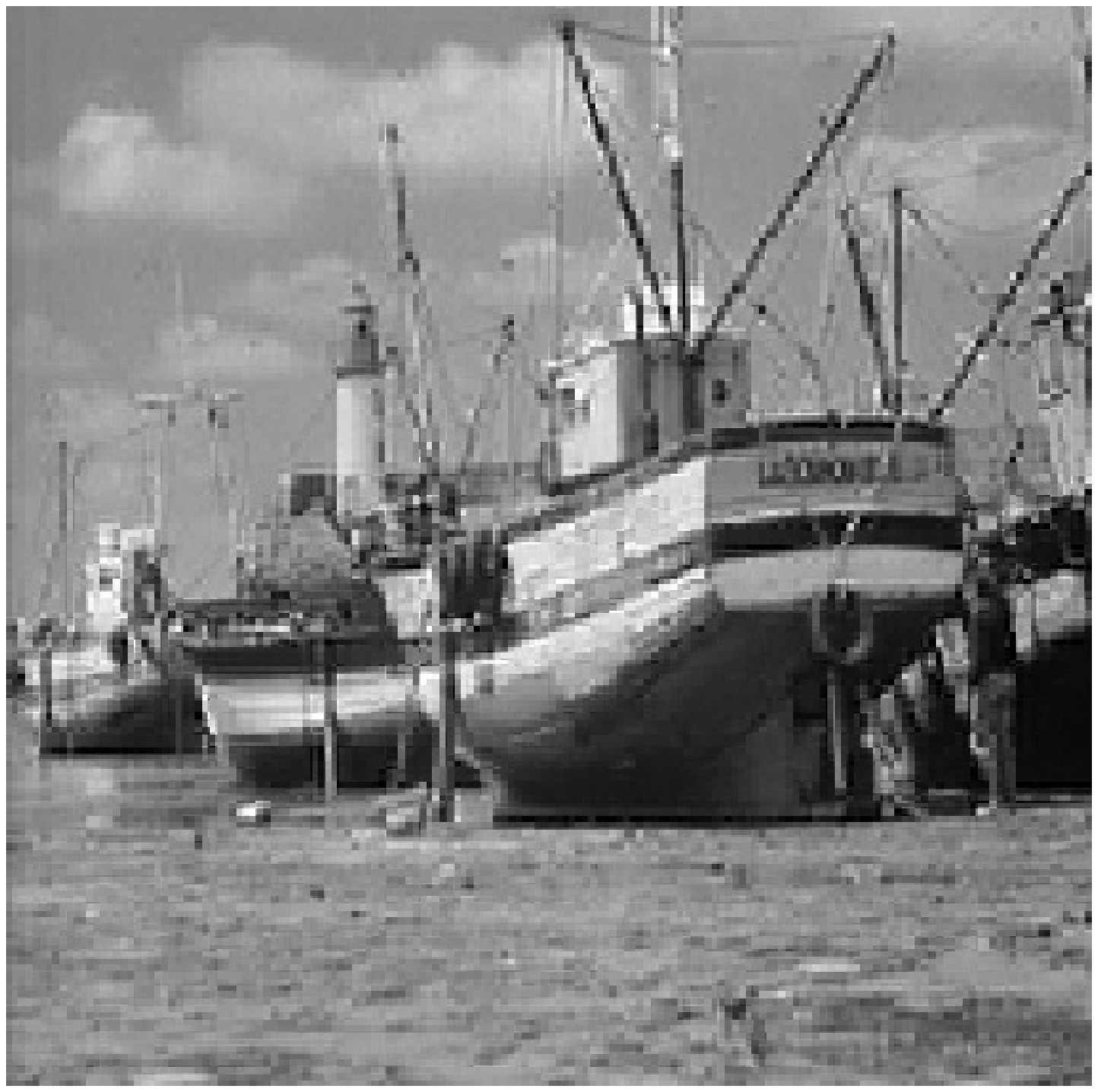} \label{boat_t0_til}}\\
\subfigure[$\tilde{\mathbf{T}}_2$~\cite{haweel2001new} (PSNR=25.760, SSIM=0.915)]
{\includegraphics[width=0.2\linewidth]{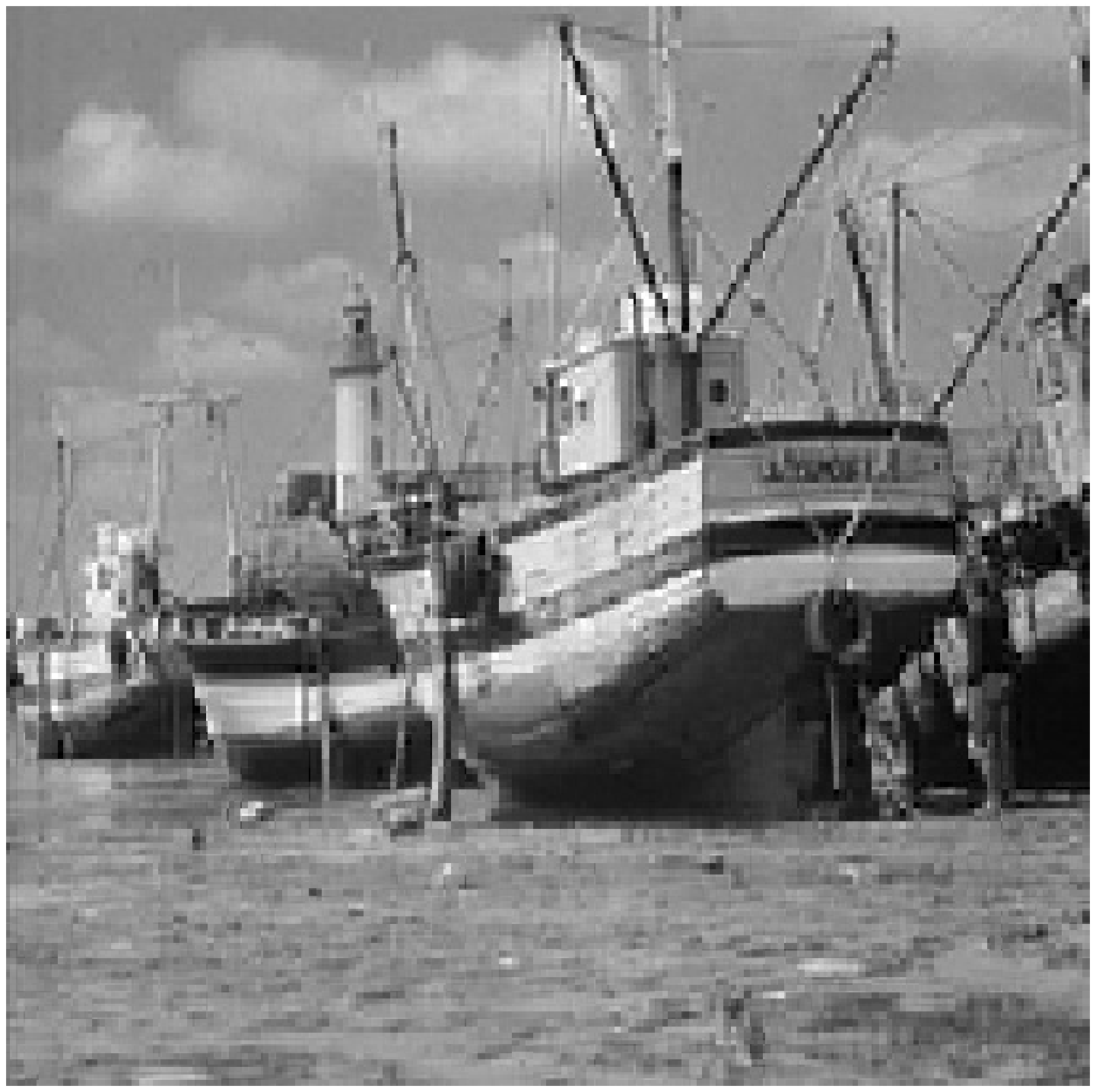} \label{boat_sdct}}
\subfigure[$\tilde{\mathbf{T}}_3$ (PSNR=28.416, SSIM=0.967)]
{\includegraphics[width=0.2\linewidth]{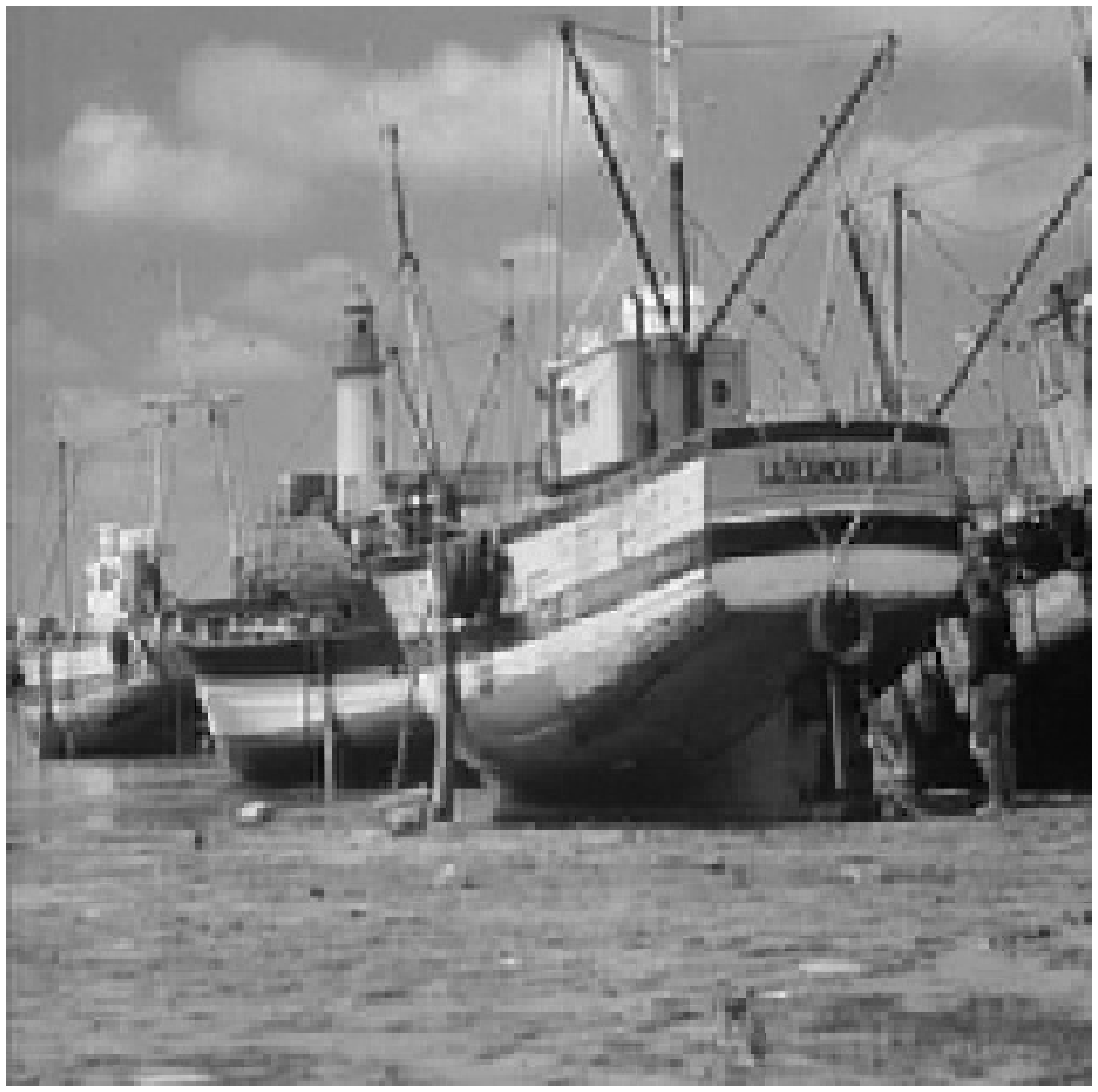} \label{boat_t2_til}}
\subfigure[DCT (PSNR=28.972, SSIM=0.970)]
{\includegraphics[width=0.2\linewidth]{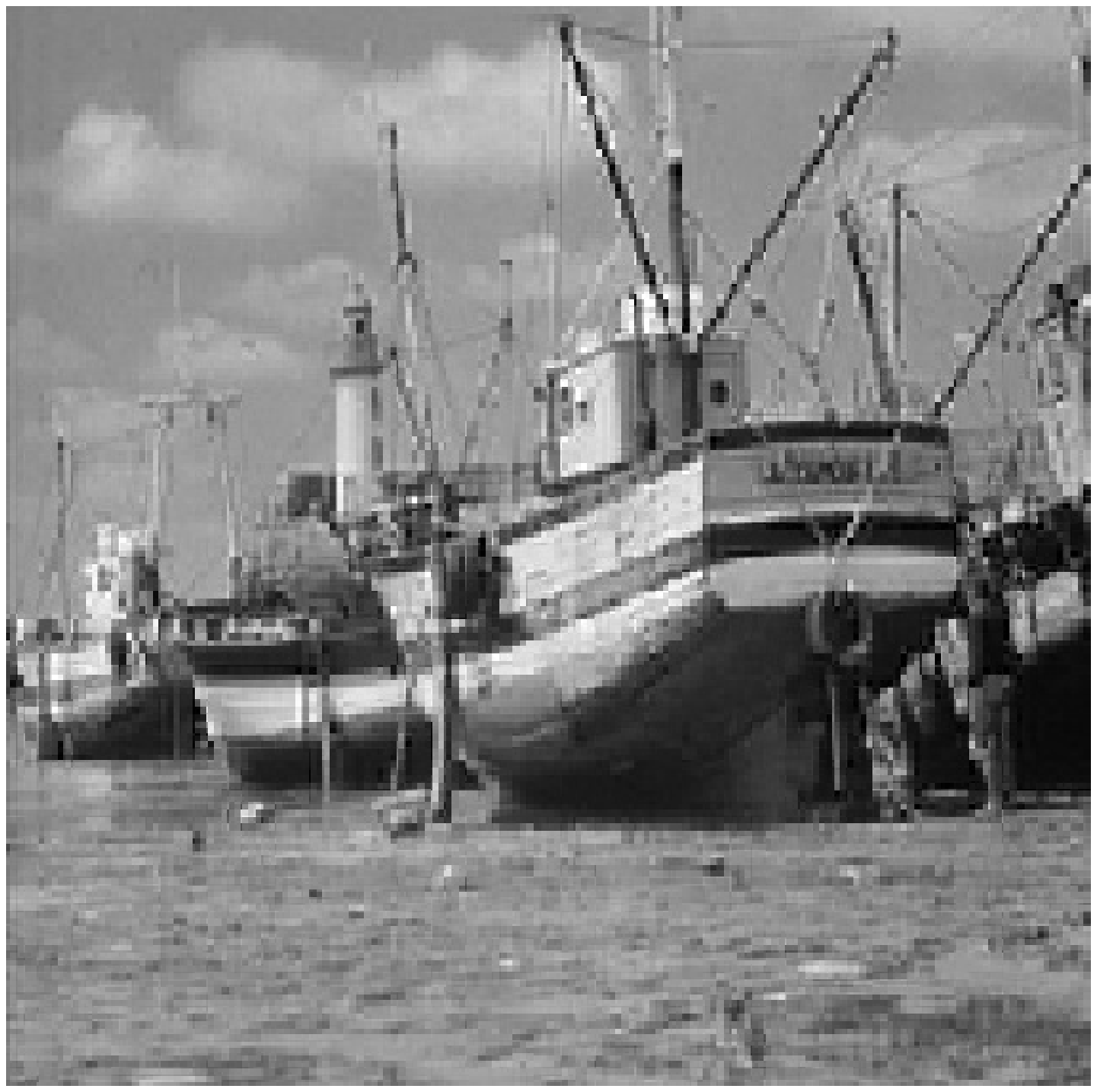} \label{boat_dct}}
\caption{
Compressed `boat' image using \subref{boat_t0} $\mathbf{T}_0$, \subref{boat_t4} $\mathbf{T}_4$~\cite{cb2011},
\subref{boat_t0_til} $\tilde{\mathbf{T}}_1$, \subref{boat_sdct} $\tilde{\mathbf{T}}_2$~\cite{haweel2001new}, \subref{boat_t2_til} $\tilde{\mathbf{T}}_3$, and \subref{boat_dct} DCT, for~$r = 10$.}
\label{fig:boat}
\end{figure*}

\begin{figure*}
\centering
\subfigure[$\mathbf{T}_0$ (PSNR=34.138, SSIM=0.989)]
{\includegraphics[width=0.2\linewidth]{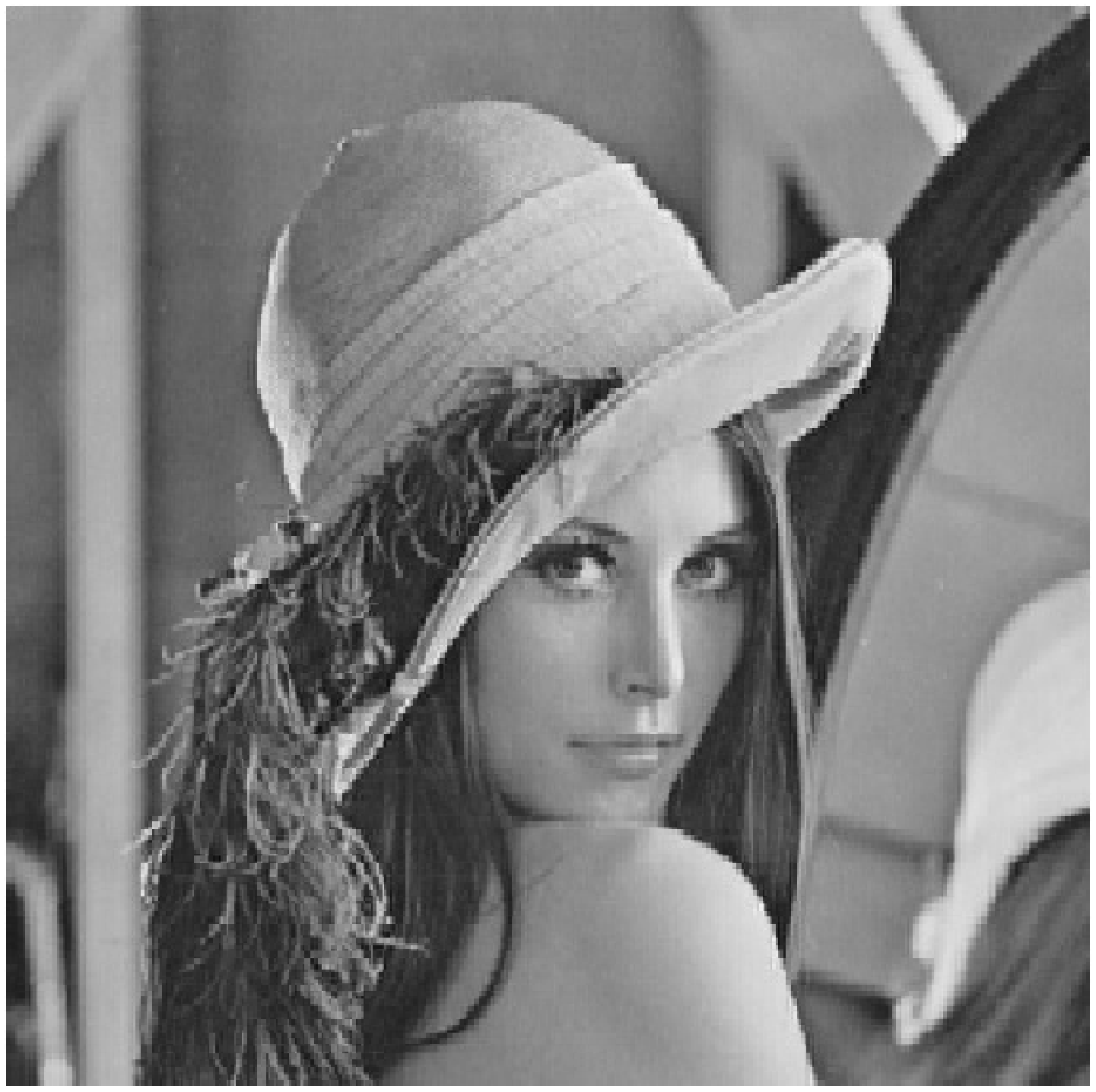} \label{lena_t0}}
\subfigure[$\mathbf{T}_4$~\cite{cb2011} (PSNR=34.159, SSIM=0.996)]
{\includegraphics[width=0.2\linewidth]{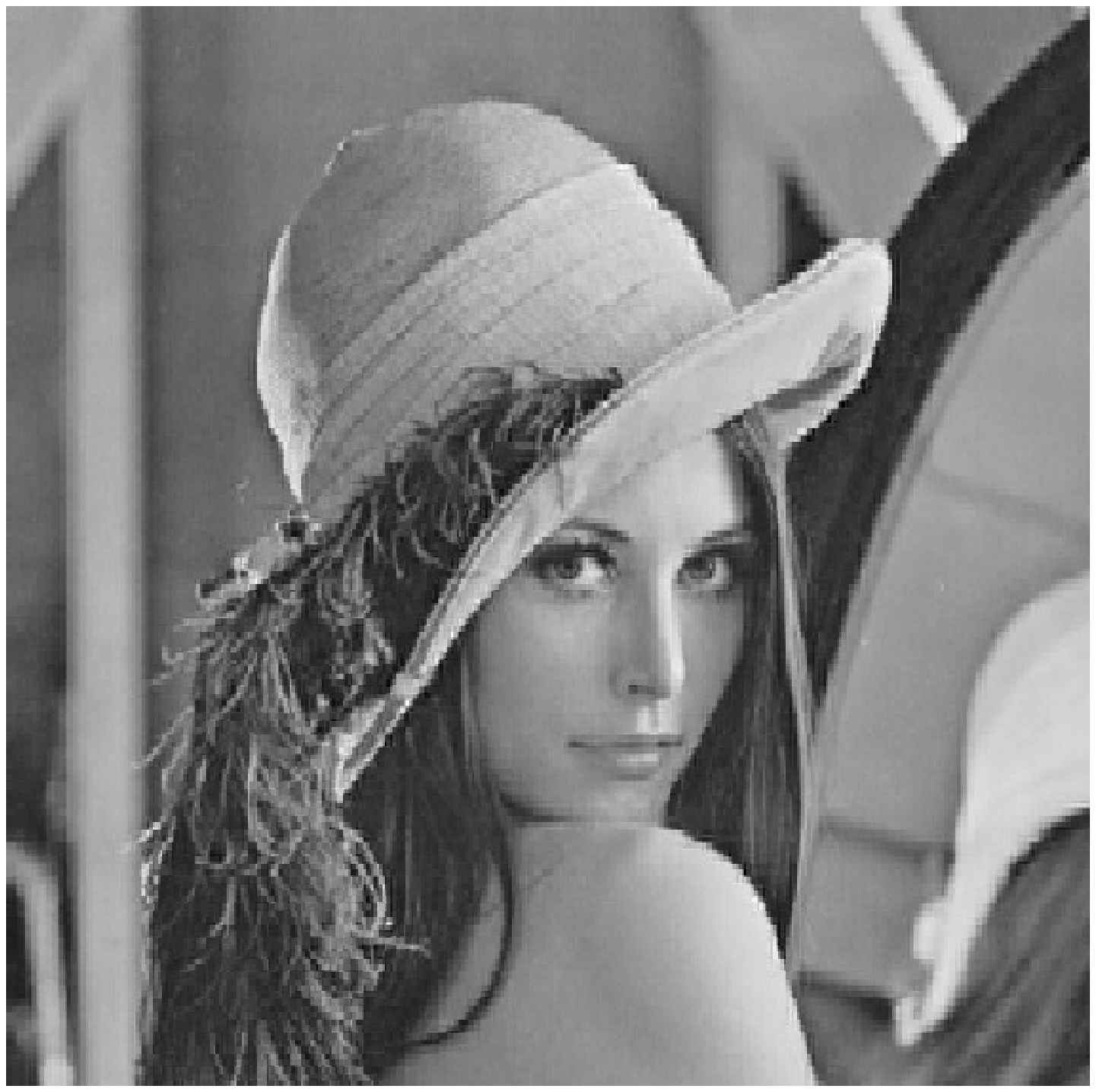} \label{lena_t4}}
\subfigure[$\tilde{\mathbf{T}}_1$ (PSNR=31.602, SSIM=0.985)]
{\includegraphics[width=0.2\linewidth]{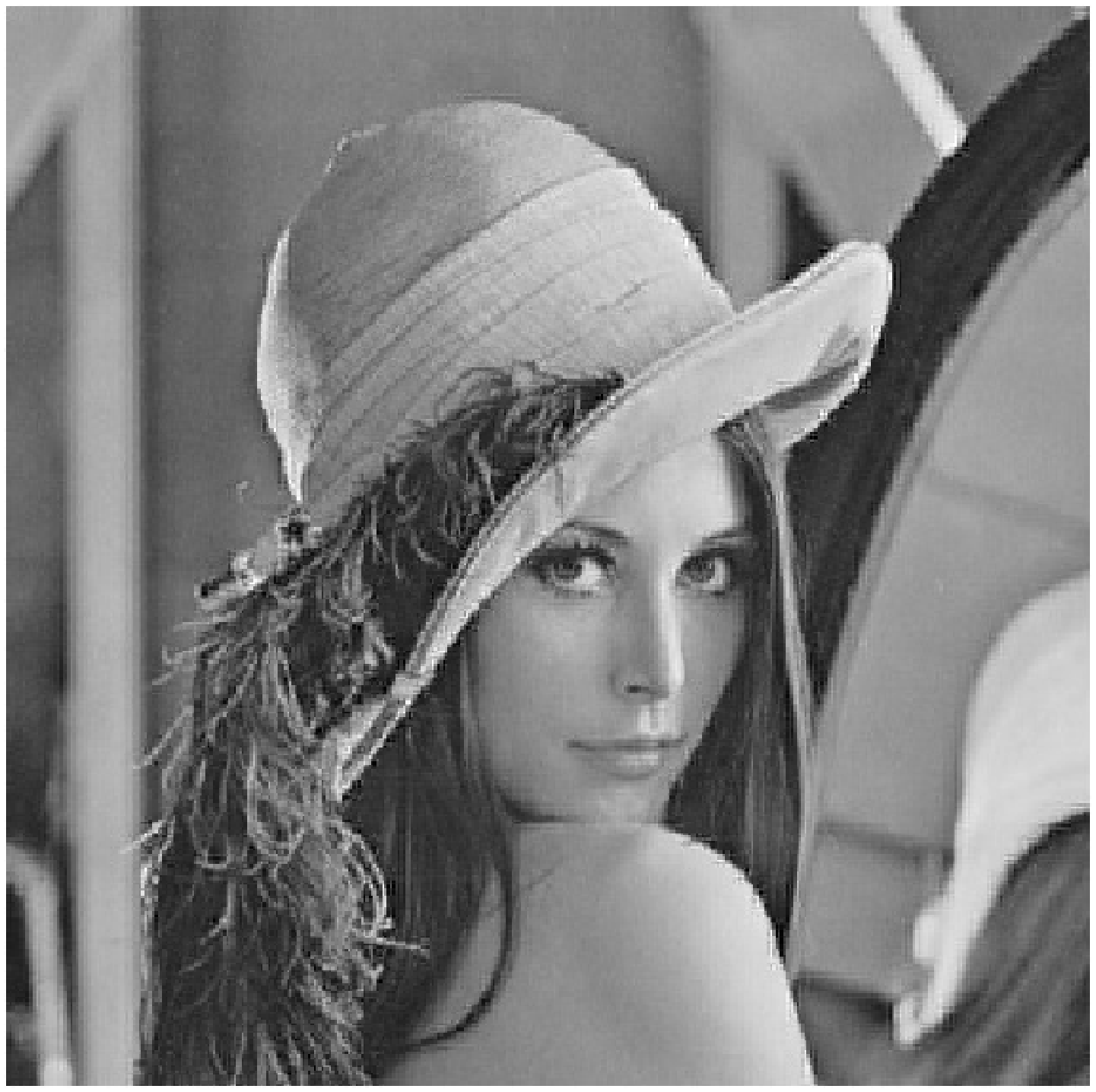} \label{lena_t0_til}}\\
\subfigure[$\tilde{\mathbf{T}}_2$~\cite{haweel2001new} (PSNR=31.852, SSIM=0.989)]
{\includegraphics[width=0.2\linewidth]{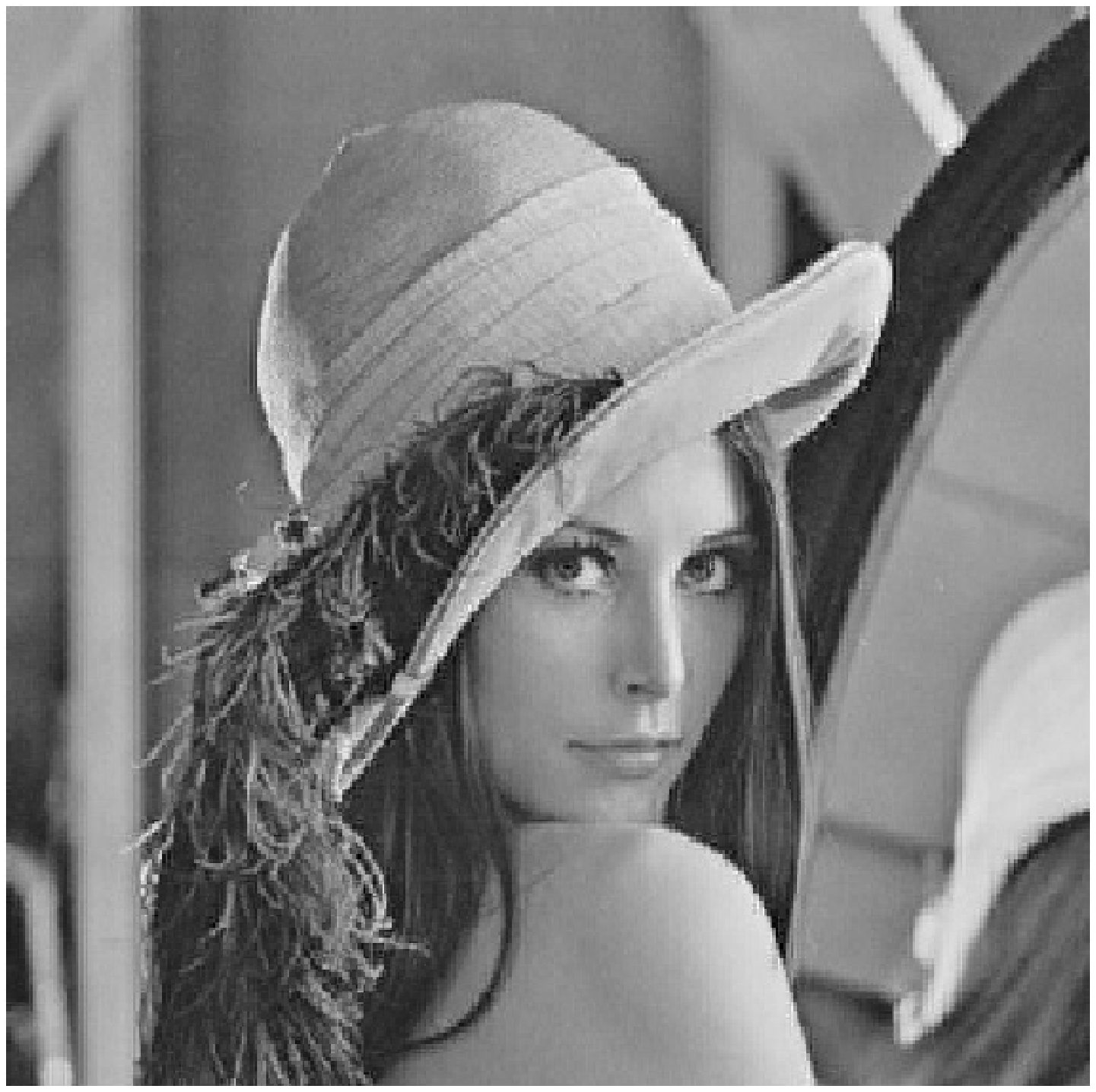} \label{lena_sdct}}
\subfigure[$\tilde{\mathbf{T}}_3$ (PSNR=35.157, SSIM=0.996)]
{\includegraphics[width=0.2\linewidth]{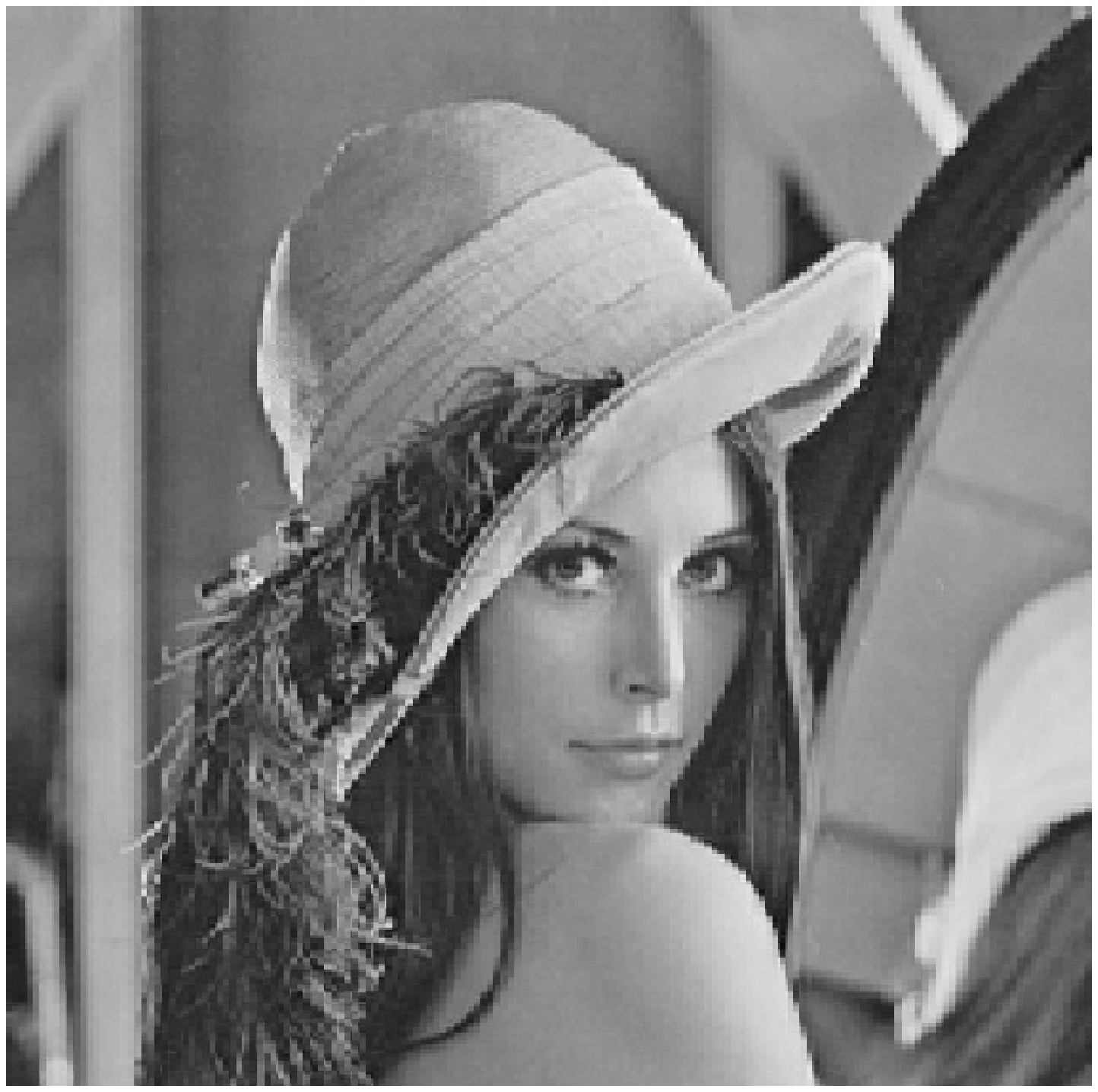} \label{lena_t2_til}}
\subfigure[DCT (PSNR=37.886, SSIM=0.997)]
{\includegraphics[width=0.2\linewidth]{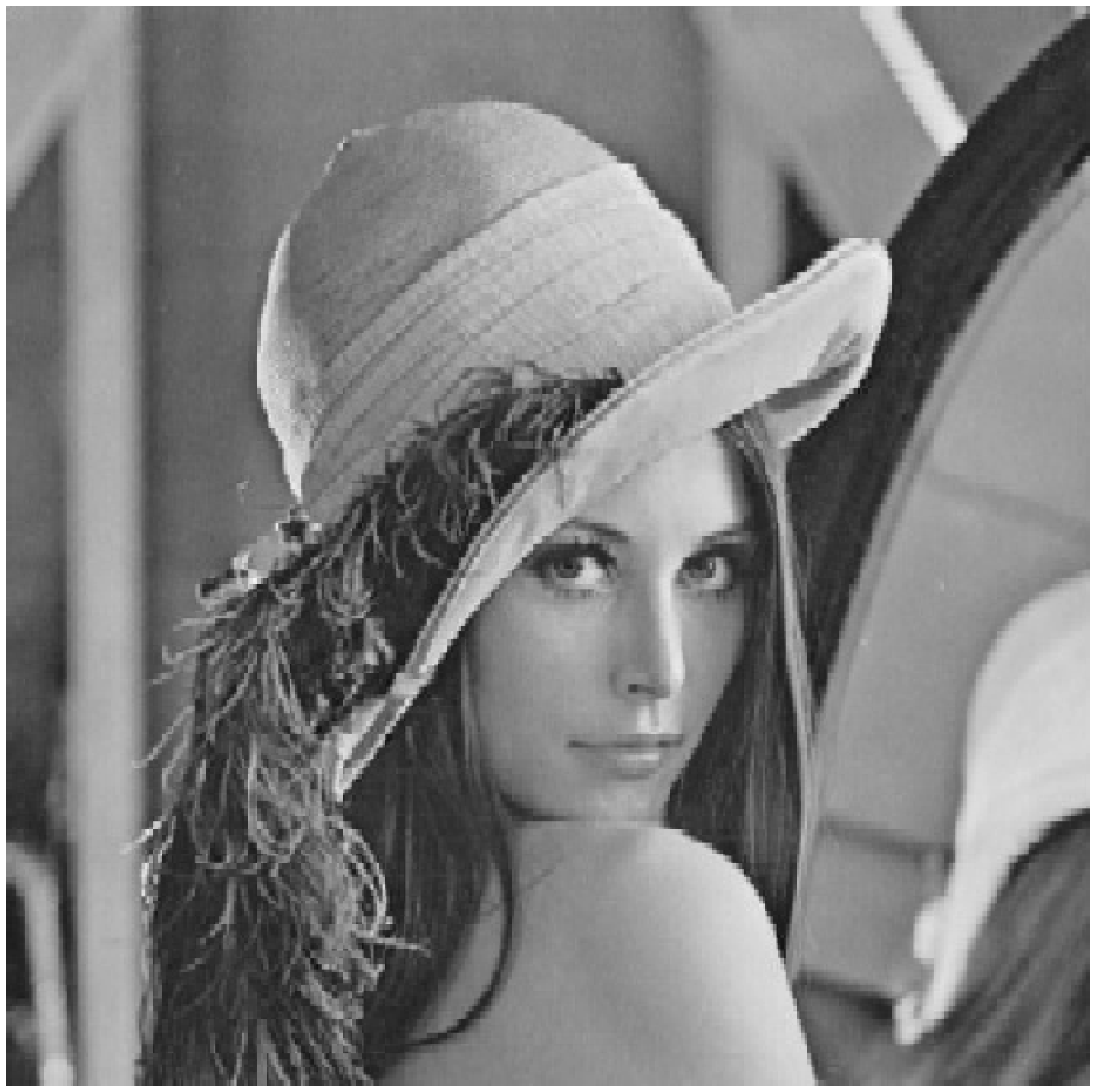} \label{lena_dct}}
\caption{
Compressed `Lena' image using \subref{lena_t0} $\mathbf{T}_0$, \subref{lena_t4} $\mathbf{T}_4$~\cite{cb2011},
\subref{lena_t0_til} $\tilde{\mathbf{T}}_1$, \subref{lena_sdct} $\tilde{\mathbf{T}}_2$~\cite{haweel2001new}, \subref{lena_t2_til} $\tilde{\mathbf{T}}_3$, and \subref{lena_dct} DCT, for~$r = 25$.}
\label{fig:lena}
\end{figure*}

The preceding discussion permit us to identify
the approximations with better 
performance and complexity trade-off.
Thus,
we separate
the following approximations:
$\tilde{\mathbf{T}}_1$,
$\tilde{\mathbf{T}}_3$,
$\mathbf{T}_0$,
and
$\mathbf{T}_4$.
Considering this restricted set of transformations,
we processed two particular images for qualitative analysis.
The SDCT and DCT were also considered for
comparison purposes. 
Fig.~\ref{fig:boat} and~\ref{fig:lena} shows 
`boat' and `Lena' images after being submitted to the JPEG-like compression experiment for $r=10$ and $r=25$, 
respectively. 
PSNR and SSIM measurements
are also included.

\section{Conclusion}
\label{sec:Conclusion}

This paper introduces a collection of DCT approximations 
derived from
the application of
common integer functions 
to the exact DCT.
The proposed mathematical formalism
could encompass---as particular cases---several transforms
already archived in literature.
In particular,
the well-known SDCT
was derived in a systematic way.
All proposed transforms
were given fast algorithms,
which have the same structure.
This suggest a common mathematical structure among all 
discussed approximations.
Only additions and simple bit-shifting operations
were necessary for their evaluation.
Such low-complexity character of the obtained
approximations
makes them suitable for hardware implementation
in dedicated architecture employing fixed-point arithmetic.
The proposed approximations
were assessed
in
terms of computational complexity and performance in JPEG-like compression;
exhibiting a good balance between cost and performance. 

\section*{Acknowledgments}

This work was partially supported by CNPq, CAPES, and FACEPE.

\footnotesize

\bibliographystyle{siam}%
\bibliography{ref-clean}

\end{document}